\documentclass[twocolumn,10pt, a4paper]{article}
\usepackage{graphicx,psfig,cite,epsf}  
\usepackage{graphics,graphicx}
\usepackage{times}
\usepackage{subfigure}
\usepackage{textcomp}
\usepackage{multirow}
\usepackage{amsmath}	
\usepackage{amssymb}	
\usepackage{gensymb}
\usepackage{siunitx}
\usepackage{url}
\usepackage{natbib}
\usepackage{hyperref}
\hypersetup{colorlinks,urlcolor=blue, citecolor=blue}
\usepackage{graphicx}
\usepackage{txfonts}
\usepackage{multirow}
\usepackage{siunitx}
\usepackage{textcomp}
\usepackage{blindtext}
\usepackage{authblk}

\providecommand{\keywords}[1]
{
  \small	
  \textbf{\textit{Keywords---}} #1
}

\def\apjl{ApJ}                
\def\apjs{ApJS}               
\def\ao{Appl.~Opt.}           
\def\aap{A\&A}                

\title{\flushleft{\bf Scientific prospects for a mini-array of ASTRI telescopes: a $\gamma$-ray TeV data challenge}}

\author[1]{F. Pintore\thanks{Corresponding author, email: fabio.pintore@inaf.it}} 
\author[1]{A. Giuliani}
\author[1]{A. Belfiore}
\author[1]{A. Paizis}
\author[1]{S. Mereghetti}
\author[1]{N. La Palombara}
\author[1]{S. Crestan}
\author[1]{L. Sidoli}
\author[2,3]{S. Lombardi}
\author[4]{A. D'A$\grave{\text{i}}$}
\author[2]{F. G. Saturni}
\author[1]{P. Caraveo}
\author[5,6]{A. Burtovoi}
\author[7,6]{M. Fiori}
\author[6]{C. Boccato}
\author[8]{A. Caccianiga}
\author[9]{A. Costa}
\author[4]{G. Cusumano}
\author[2]{S. Gallozzi}
\author[6]{L. Zampieri}
\author[10]{B. Balmaverde}
\author[ ]{ for the ASTRI Project \thanks{\protect\url{http://www.brera.inaf.it/\~astri/wordpress/}}}
\author[11]{L. Tibaldo}

\affil[1]{\footnotesize INAF-IASF Milano, via A. Corti 12, I-20133 Milano, Italy. }
\affil[2]{ \footnotesize ASI Space Science Data Center, 00133 Roma, Italy.}
\affil[3]{ \footnotesize INAF Osservatorio Astronomico di Roma, 00078 Monte Porzio Catone, Roma, Italy }
\affil[4]{\footnotesize  INAF-IASF Palermo, 90146 Palermo, Italy.}
\affil[5]{\footnotesize  Centre of Studies and Activities for Space (CISAS) ``G. Colombo'', University of Padova, Via Venezia 15, 35131 Padova, Italy. }
\affil[6]{\footnotesize  INAF-Osservatorio Astronomico di Padova, Vicolo dell'Osservatorio 5, I-35122 Padova, Italy. }
\affil[7]{ \footnotesize Department of Physics and Astronomy, University of Padova, Via F. Marzolo 8, 35131, Padova, Italy. }
\affil[8]{\footnotesize  INAF-Osservatorio Astronomico di Brera, via Brera 28, 20121 Milano, Italy. }
\affil[9]{\footnotesize  INAF-Osservatorio Astronomico di Catania, 95123 Catania, Italy. }
\affil[10]{ \footnotesize INAF-Osservatorio Astrofisico di Torino, Via Osservatorio 20, I-10025 Pino Torinese, Italy.}
\affil[11]{\footnotesize  Institut de Recherche en Astrophysique et Plant\'eologie, Universit\'e de Toulouse, CNRS, CNES, UPS, 9 avenue Colonel Roche, 31028 Toulouse, Cedex 4, France. }

\begin{document}

\pagestyle{plain}
\pagenumbering{arabic}

\twocolumn[
  \begin{@twocolumnfalse}
    \maketitle
    \begin{abstract}
{\it ASTRI} is a project aiming at the realization of a $\gamma$-ray imaging Cherenkov telescope that observes the sky {in the TeV band}. Recently, the development of a mini-array (MA) of ASTRI  telescopes has been funded by the Istituto Nazionale di Astrofisica.
The ASTRI Comprehensive Data Challenge (ACDC) project aims at optimizing the scientific exploitation and analysis techniques of the ASTRI MA, by performing a complete end-to-end simulation of a tentative scientific program, from the generation of suitable instrument response functions to the proposal, selection, analysis, and interpretation of the simulated data.
We assumed that the MA will comprise nine ASTRI telescopes arranged in a (almost) square geometry (mean distance between telescopes of $\sim$250m).
We simulated three years of observations, adopting a realistic pointing plan that takes into account, for each field, visibility constraints for {an assumed} site in {Paranal (Chile)} and observational time slots in dark sky conditions. We simulated the observations of nineteen Galactic and extragalactic fields selected for their scientific interest, including several classes of objects (such as pulsar wind nebulae, supernova remnants, $\gamma$-ray binaries etc), for a total of {81} point-like and extended sources.
Here we present an overview of the ACDC project, providing details on the different {software} packages needed to carry out the simulated three-years {operation} of the ASTRI MA. We discuss the results of a systematic analysis applied on the whole simulated data, by making use of {prototype} science tools widely adopted by the TeV astronomical community. Furthermore, particular emphasis is also given to some targets used as benchmarks.
    \end{abstract}
 
  \end{@twocolumnfalse}
]

\section{Introduction}

In the last years, substantial progress has been achieved in the field of $\gamma$-ray astronomy thanks to the new generation of ground-based telescopes based on  the imaging atmospheric Cherenkov technique (IACT, see e.g. \citealt{deNaurois15}, for a review). 
More than 200 {very-high-energy} sources are now known (\url{http://tevcat.uchicago.edu/}),   
belonging  to different classes. Amongst Galactic sources, the most numerous classes are pulsar wind nebulae (PWN) and supernova remnants (SNR), but also some pulsars (Crab and Vela) and some binary systems have been detected. Extragalactic sources are typically blazars plus a minority of other types of active galactic nuclei (AGN) and {starburst galaxies}. Finally, a significant fraction of TeV sources are still unidentified \citep[e.g.][]{HESScollaboration+18b}. 

The next large facility to further advance in this field is the Cherenkov Telescope Array (CTA, \citealt{cta19}). CTA will cover the {energy} range from 20 GeV up to 300 TeV with arrays of telescopes of different sizes, placed at two different sites in order to cover both the southern and northern celestial hemispheres\footnote{https://www.cta-observatory.org/}. CTA will provide a sensitivity {about} one order of magnitude better than that of current {IACT} telescopes.
As of 2019, CTA-North will comprise two classes of telescopes with different dimensions: Large Size Telescopes (LST) of 23 m diameter (required energy range of 0.02--3 TeV) and  Medium Size Telescopes (MST) of {about} 12 m diameter (required energy range of 0.08--50 TeV). The CTA-South array, in addition to LST and MST, will include a significant number of Small Size Telescopes (SST) of $\sim$4 m diameter (required energy range of 1--300 TeV), in order to provide a good sensitivity also at the highest energies, which are mostly relevant for sources in our Galaxy. 

\begin{table*}[ht]
  \begin{center}
\footnotesize
   \caption{Selected targets for the {data challenge} simulations.} 
      \label{src_tab}
   \begin{tabular}{|c l l c | l | l | c |}
\hline 
Field & \multicolumn{1}{c}{RA} & \multicolumn{1}{c}{Dec} & Exposure & \multicolumn{1}{c|}{Main sources} & \multicolumn{1}{c|}{Type} &  Extended \\
 & \multicolumn{1}{c}{deg} & \multicolumn{1}{c}{deg} & \multicolumn{1}{c|}{hr} & & & \\
\hline
\multirow{2}{*}{01+02$^{**}$} & \multirow{2}{*}{276.63} & \multirow{2}{*}{$-13.09$} & \multirow{2}{*}{50+250} & {\bf LS 5039}$^*$ & Binary & n \\
 & & & & {\bf HESS J1825-137$^*$} & PWN & y \\
 \hline
03 & 40.67 & -0.01 & 200 & NGC 1068 & AGN & n  \\
 \hline
04 & 15.04 & -33.71 & 100 & Sculptor & dSph galaxy & n \\
 \hline
05 & 53.93 & -54.05 & 100 & Reticulum II & dSph galaxy & n \\
 \hline
06 & 342.98 & -58.57 & 100 & Tucana II & dSph galaxy & n \\
 \hline
07 & 237.25 & -30.75 & 300 & Te-REX 1549 & AGN & n \\
 \hline
\multirow{2}{*}{08} & \multirow{2}{*}{266.92} & \multirow{2}{*}{$-26.47$} & \multirow{2}{*}{300} & HESS J1748-248$^*$ (Ter 5) & Globular cluster & n \\
 & & & & G0.9+0.1 & PWN & y \\
 \hline
09 & 195.75 & -63.20 & 300 & HESS J1303-631 & PWN & y \\
 \hline
10 & 84.00 & -67.59 & 200 & {\bf LMC P3} & Binary & n \\
 \hline
\multirow{2}{*}{11} & \multirow{2}{*}{248.25} & \multirow{2}{*}{$   -47.55$} & \multirow{2}{*}{150} & HESS J1632-478$^*$ & PWN? & y \\
 & & & & HESS J1634-472$^*$ & unknown & y \\
 \hline
\multirow{2}{*}{12} & \multirow{2}{*}{269.72} & \multirow{2}{*}{$-24.05$} & \multirow{2}{*}{300} & {\bf SNR W28}$^*$ & SNR & y \\
 & & & & HESS J1748-248 (Ter 5) & Globular cluster & n \\
 \hline
13 & 155.99 & -57.76 & 200 & Westerlund 2 & Star cluster & y \\
\hline 
14 & 128.75 & -45.60 & 100 & Vela & PSR and PWN & y \\
 \hline
\multirow{2}{*}{15} & \multirow{2}{*}{266.85} & \multirow{2}{*}{$-28.15$} & \multirow{2}{*}{400} & G0.9+0.1$^*$ & PWN & y \\
 & & & & HESS J1748-248 (Ter 5) & Globular cluster & n \\
 \hline
16 & 228.53 & -59.16 & 200 & MSH 15-52 & PWN & y \\
 \hline
\multirow{3}{*}{17} & \multirow{3}{*}{278.39} & \multirow{3}{*}{$-10.57$} & \multirow{3}{*}{300} & HESS J1833-105$^*$ & PSR and PWN & n \\
 & & & & HESS J1825-137 & PWN & y \\
 & & & & LS 5039 & Binary & n \\
 \hline
18 & 83.63 & 22.01 & 100 & {\bf Crab} & PSR and PWN & n  \\
 \hline
19 & 329.72 & -30.22 & 200 & PKS 2155-304 & AGN & n \\
 \hline
\end{tabular}
\flushleft $^*$ Reference target(s) in the field. We highlight in bold the sources discussed in detail in the paper.\\
$^{**}$ Same fields but corresponding to the high and low flux state of LS 5039.
\end{center}
\end{table*}

The {\it Horn-D'Arturo} ASTRI\footnote{Astrofisica con Specchi a Tecnologia replicante Italiana, \url{http://www.brera.inaf.it/~astri/wordpress}} telescope \citep{pareschi16,maccarone17,scuderi18} is {one of the prototypes for} the SST based on a dual-mirror Schwarzschild-Couder optical design \citep{vassiliev07,canestrari13}, that has been developed in the context of the Italian participation to the CTA project. The diameter of the primary and secondary mirrors of the telescope are 4.3 meters and 1.8 meters, respectively. The design is such that the telescope benefits from a large field of view (FoV) of $\sim10$\textdegree\ \citep{rodeghiero16}. The ASTRI camera is based on very-fast response SiPM sensors \citep[of about ten ns][]{catalano18}, which provide high detection efficiency of photons and high {single photo-electron resolution} \citep[e.g.][]{bonanno16,romeo18}.  
The ASTRI prototype is currently operating at the Serra La Nave observing station on Mount Etna in Sicily \citep{maccarone13} and very recently has obtained the detection of the Crab nebula at TeV energies \citep{lombardi19}.

Based on the ASTRI prototype, a mini-array composed of nine ASTRI telescopes is being developed and {led} by Istituto Nazionale di Astrofisica (INAF). The design of the ASTRI telescopes will permit observations at very high energies (up to $\sim300$ TeV) with a large FoV of $\sim$$10$\textdegree\ diameter. Therefore the mini-array will be able to observe several $\gamma$-ray sources simultaneously. 
Initially, the mini-array was intended to be developed in the context of the preparatory effort for participating in CTA-South, at the Paranal (Chile) site. However, on June 2019, the mini-array became a project independent of CTA and it was decided to locate it in the Northern hemisphere, at Mount Teide in the Canary island of Tenerife\footnote{{ At the beginning of the ACDC project, the mini-array was still considered as a part of the future CTA-South. For this reason, we carried-on the whole project assuming the location of the mini-array at the Paranal site, and not the Teide site. 
}}. It is expected that the mini-array will begin operations within 2023.

Here, we discuss the {\it ASTRI Comprehensive Data Challenge} (ACDC) project that aims at characterizing the capabilities of an ASTRI mini-array and investigating the scientific exploitation of this innovative instrument in the {era of the current generation of Cherenkov telescopes and CTA}. The ACDC was not intended to provide observational constraints to the ASTRI mini-array but rather to test our end-to-end data analysis chain.
Therefore, the results presented hereafter are to be considered examples of the potential instrument's capabilities, independently on the final ASTRI mini-array site. Obviously, the data challenge will be repeated for the actual location using the updated array geometry on a different sample of sources which will partially overlap with the one used here.

\section{The ACDC Project}

\subsection{Simulating three years of ASTRI mini-array observations}

\begin{figure*}
   \centering
   \includegraphics[width=17cm]{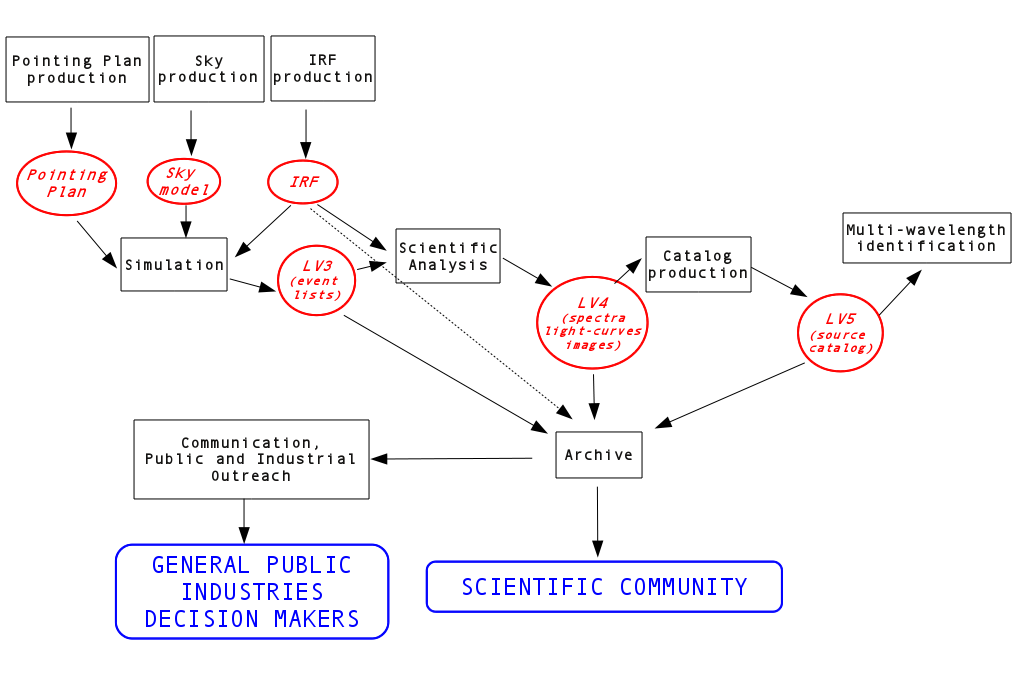}
   \includegraphics[width=\hsize]{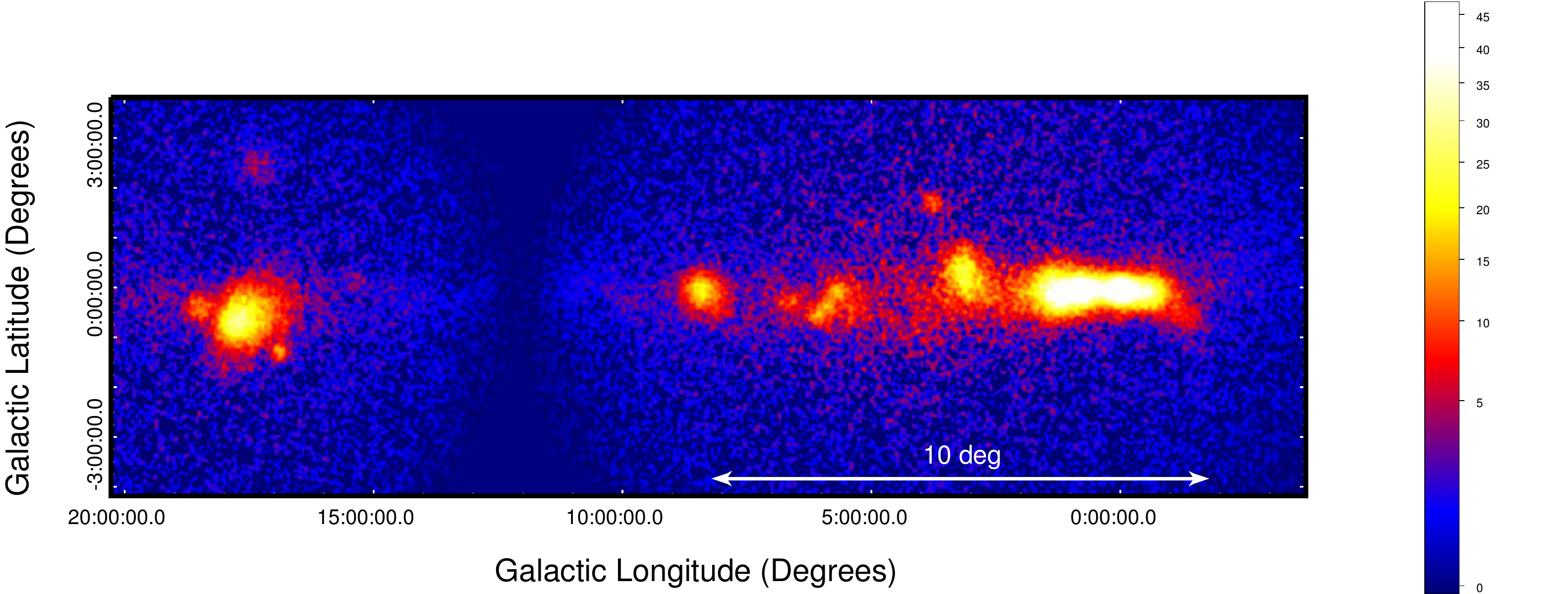}
      \caption{Top: work flow of the ACDC project. Bottom: simulation of the Galactic plane. The possibility to observe simultaneously several sources can be appreciated thanks to the large FoV ($\sim10$\textdegree\ diameter). The map (in Galactic coordinates) was created for a total of ~2250 pointings of ~20 minutes each. Note that this map is background subtracted, not corrected for exposure and the color-bar is in counts per pixel (pixel size of 0.02 deg). {The diameter of the ASTRI FoV is indicated with the white arrow.}}
         \label{plane}
\end{figure*}

The aim of ACDC is to perform a simulation of the observations of {TeV gamma-rays} representative of the data we expect to obtain with the ASTRI mini-array {\bf and to test the end-to-end analysis chain}.
In order to obtain a realistic three-year long set of observations\footnote{Here and in the following, when we refer to ``observations'', we clearly mean simulated observations.}, we performed a sequence of activities and simulated operations like those expected to occur in reality. The {ACDC} project was divided in different work packages (WP, see the scheme in Figure~\ref{plane}) and hereafter we describe the ones relevant to this work.

We started by issuing a ``Call for Proposal'' in order to select the potential sources of interest in the southern-sky visible in the hypothesis of a mini-array of SSTs in Paranal using the ASTRI design (see above for caveats on this scenario). Then we proceeded with all the activities needed to simulate a complete end-to-end data-challenge: planning of the observations, generation of the instrument response function (IRF) of the system, production and archiving of the simulated data, analysis of the data and production of a catalog,  multi-wavelength association of the detected sources, and,  finally the production of a final archive of all scientific products.

\subsection{Selection of the sample}

{The call for proposals, open to all {ASTRI members}, was for observations covering three years of  operations. The accepted proposals span a variety of scientific cases, from dark matter candidates to acceleration of particles by neutron stars. They include the} 20 {reference} targets (in 18 fields) listed in Table~\ref{src_tab}, corresponding to a total requested time of 4350 hours. Although these  targets represent only a partial view of the southern sky, they were selected because of their high scientific interest. They belong to the classes of pulsar wind nebulae (PWN), $\gamma$-ray binaries, active galactic nuclei (AGN), dwarf spheroidal galaxies (dSph galaxies) dark matter candidates and supernova-remnants (SNR).

\subsection{Pointing plan}

One task of ACDC ({\it Pointing Plan Production} WP) was devoted to the production of a realistic observation schedule, matched on the time period from January 1, 2020 to January 1, 2023\footnote{considered for simplicity and not optimized for possible science windows}, taking into account the visibility constraints and Moon conditions at {Paranal}. We estimated that this corresponds to $\sim$5160 hr of useful time in the three years. {This is longer than the exposure time requested in the call for proposal (4350 hr) for the reference targets. In order to exploit all the available time, we used the time in excess ($\sim$810 hr) to simulate blank sky fields (see below) and to increase, when possible, the exposure of the weak and extended reference targets.}

We produced a run-schedule that optimised the temporal allocation, the observation {elevation angle} (chosen to be $>55$ deg {above the horizon}) and that minimised the number of slews between different fields. {We took into account only the temporal windows not disturbed by Moon light (i.e. when the Moon is below horizon or new Moon epochs). We did not include any time loss due to weather conditions.}
All the observations were divided in blocks of $\sim20$ minutes following a {dithering pattern as follows}: the pointing direction of each block was chosen randomly from a uniform distribution of directions within a radius of one degree from the reference target coordinates. This choice describes a realistic schedule where observing conditions and pointing directions may change not in a regular way. Therefore, each observation consists of a mosaic of blocks in which the sources are at different positions in the FoV. The adopted observational approach has the advantage of further enlarging the size of the ASTRI FoV, allowing us to smooth any structure of the instrumental background and to observe sources that would have been at the edge of the FoV in case of fixed pointing. 
In fact, a traditional wobbling observing mode is not {applied} in the case of our observations characterized by a large field of view containing several targets. 

\begin{figure*}
   \centering
   \includegraphics[width=8.8cm]{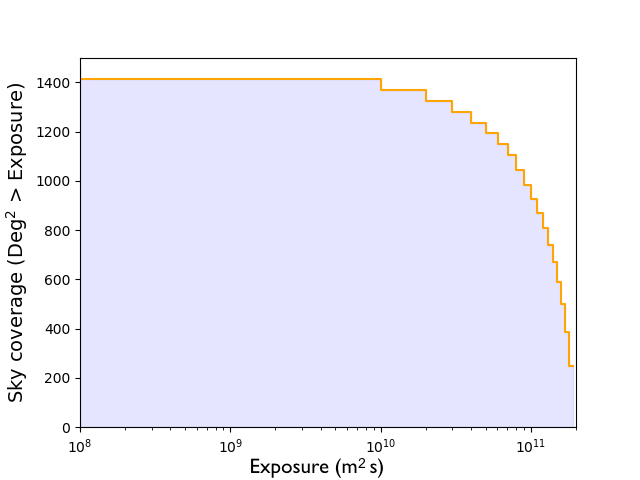}
   \includegraphics[width=8.8cm]{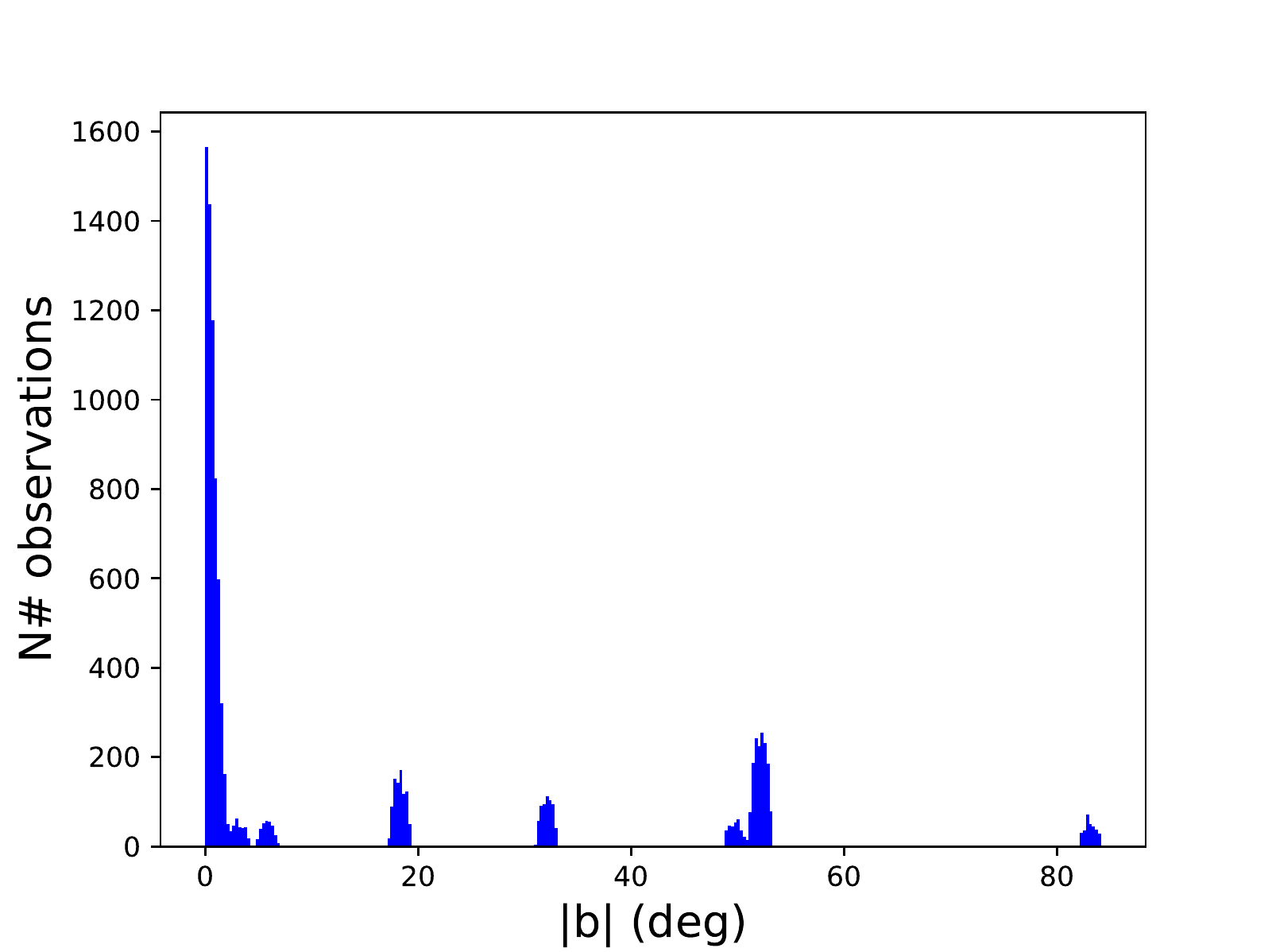}
      \caption{Left: integral distribution, in the 2.5--10 TeV energy range, of the sky coverage (in squared degree) as a function of the exposure (defined as the effective area times the net observation time). Right: number of simulated observations in terms of the (module of) Galactic latitude.}
         \label{fields}
\end{figure*}

\subsection{Background simulation}

In each observation we also included the background contribution, given by the combination of the diffuse Galactic $\gamma$-ray background and {the cosmic background} ({\it Sky Production} WP). 
The diffuse Galactic background is due to the $\gamma$-ray emission from the interstellar medium, and it is based on predictions from codes solving the cosmic-ray (CR) transport equations and calculating the related multi-wavelength emission. The models used are informed by observations including direct CR measurements and high-energy $\gamma$-ray observations by Fermi-LAT. Inverse-Compton (IC) emission is produced by interactions of CR electrons and positrons on low-energy photons. We used predictions from {\it Picard} that assume a model with {four} spiral arms for the CR source distribution, which has large imprint on the resulting IC emission \citep{werner15}.
Emission from interstellar gas is produced by CR nuclei hadronic interactions and by electron/positron Bremsstrahlung. We use predictions from {\it Dragon}, that assume position-dependent diffusion and convection properties to reproduce the intensity/hardening of interstellar gamma-ray emission seen by Fermi-LAT toward the inner Galaxy \citep{gaggero15}.

The residual {non-$\gamma$} background was estimated through dedicated Monte Carlo simulations of cosmic ray and electron events.
It was included into the IRF (see next Section) as a given spatial model rate, dependent only on the energy and the radial positions {in detector coordinates} on the camera, assuming radial symmetry. 
We note that about 800 hr of the pointing plan were devoted to observations of sky fields not containing TeV sources. These ``blank'' fields permit to perform ON-OFF analyses {and} to estimate the level of background when analyzing very extended sources or sources located in crowded fields.

\subsection{The simulation of the $\gamma$-ray sky}

Since the ASTRI telescope has a good sensitivity up to off axis angles of $\sim$3--4 degrees, several sources {can} be contained in the FoV, as shown in the example of Fig.~\ref{plane}. Therefore, besides the selected targets, we also included in our sky model all the known $\gamma$-ray sources falling in the FoV, even considering the increased sky regions covered thanks to the dithering pattern around the nominal target direction. The spectral and spatial  properties of all the simulated sources were based on the results reported in the H.E.S.S.\footnote{https://www.mpi-hd.mpg.de/hfm/HESS/pages/home/sources/} and Fermi catalogues \citep{3FGL}.
Finally, in addition to the known sources, we included eight (six transient and two persistent) sources reaching a peak flux comprised in the range $\sim$0.05-1$\%$ Crab units (in the 2.5--95 TeV energy band), in order to test the possibility of serendipitous Galactic or extragalactic discoveries of new sources. For these targets, we considered power-law or cut-off power-law spectral models, with {random spectral parameters (photon indexes in the range [1.4--2.7])}, and we also considered a random temporal variability as outbursts, flares or persistent emission. 
Therefore, we simulated a total of 81 sources, the vast majority being spatially extended and Galactic in nature. They comprise PWN (twenty-nine objects), SNR (nine objects), binary systems (three objects), AGN or galaxies (seven objects) with the rest being unidentified sources. The spectral shapes of the targets are, in most cases, modelled with power-laws ($\Gamma$ in the range 1.4--4.5 and normalization at 3 TeV between $\sim10^{-14}-3\times10^{-12}$ photons cm$^{-2}$ s$^{-1}$ TeV$^{-1}$). In a few cases either  exponentially cut-off power-laws (with cut-off between 4-11 TeV) or log-parabola models were used. For more complex models for which an analytical description was not possible, we used tabulated models (three cases). 
We modelled the morphology of spatially extended sources with {2-D radial Gaussians (with symmetric standard deviation $\sigma$}, in the range $\sim$0.1--2 deg), {discs and ellipses} (size 0.1--0.2 deg), or, in case of more complex shapes, we used energy-dependent template maps. The latter were generated in accordance to the source spatial and spectral properties taken from the literature (when available).

The whole observation program resulted in the coverage of a total of about 1250 square degrees of sky. The integral distribution of the exposure is shown in  Fig.~\ref{fields} (left panel). In the right panel of Fig.~\ref{fields}, we show the distribution of the sky exposure as a function of Galactic latitude. It is clear that most of the exposure was devoted to the Galactic plane. 

\subsection{Simulation of the Sky, data storage and multi-wavelength associations}

 \begin{figure*}
   \centering
   \includegraphics[width=8.8cm]{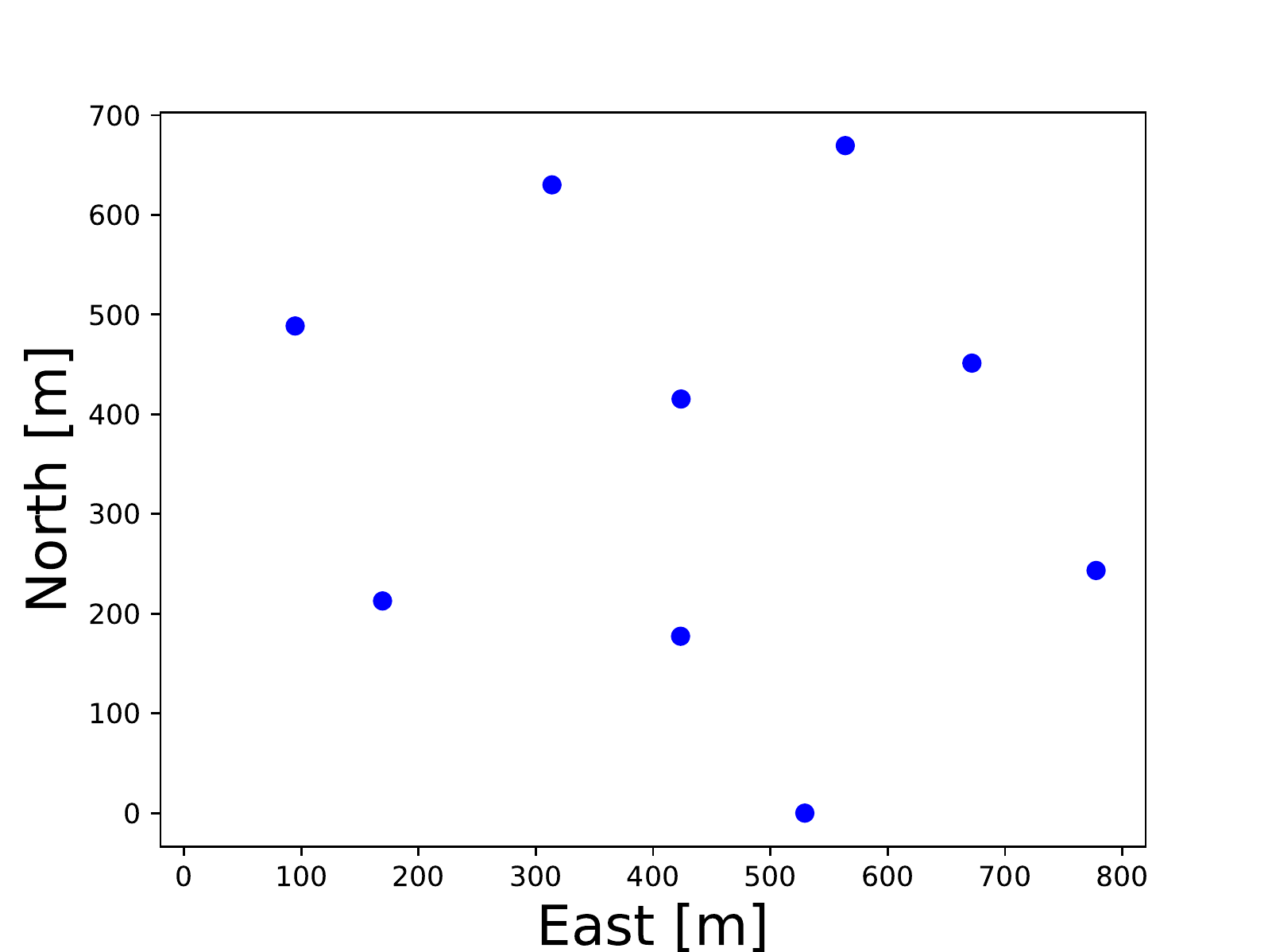}
   \includegraphics[width=9.8cm]{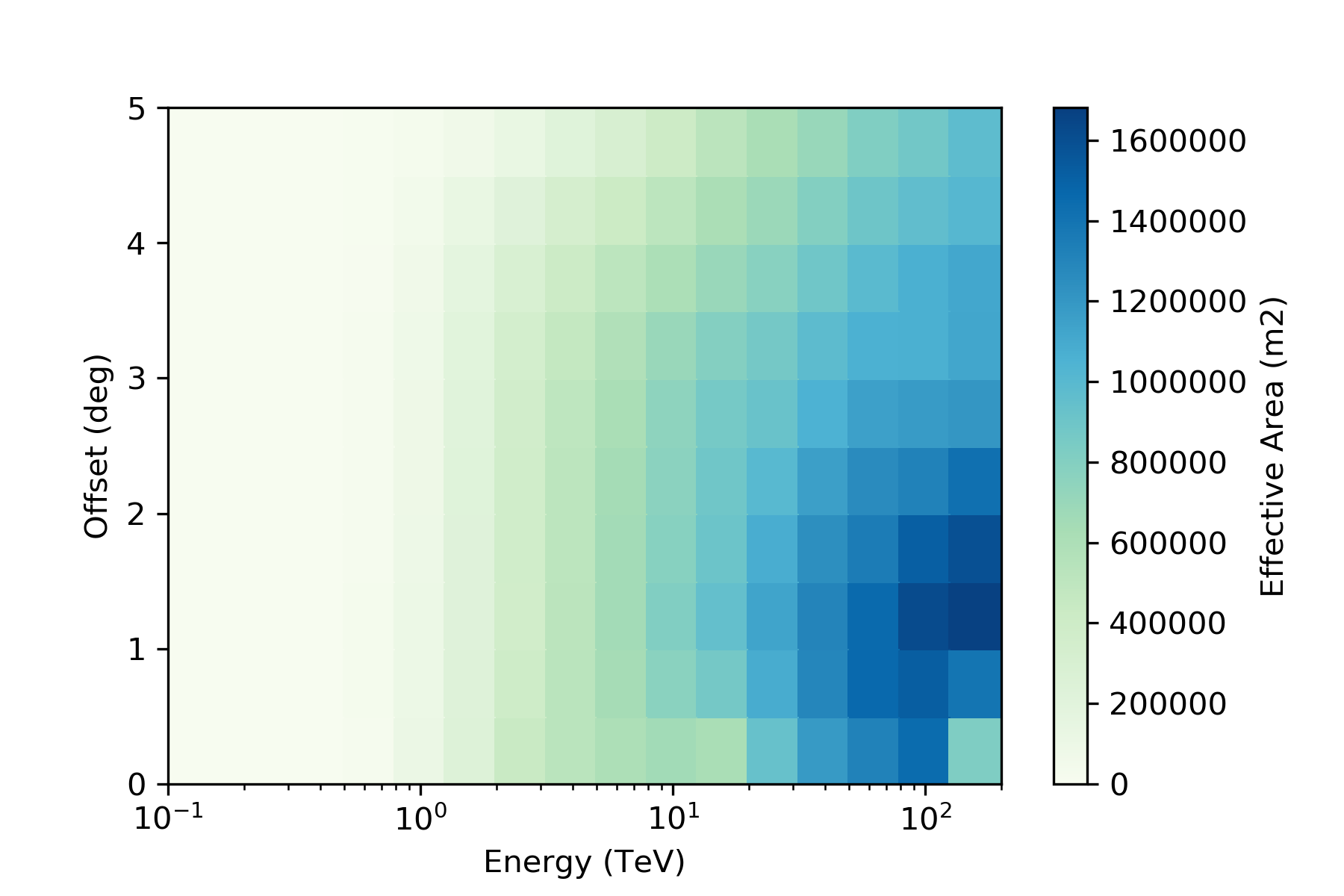}
       \caption{Left: the blue points represent the positions of the ASTRI telescopes in the mini-array layout considered in this work, which is compliant with one of the possible configurations of a sub-array of SSTs at the CTA Southern site.  Right:effective area of the ASTRI mini-array as a function of energy and offset angle.}
          \label{layout_ASTRI}
 \end{figure*}

In the {\it Simulation} WP, we used the tool {\sc ctobssim} of {{\sc ctools}-1.5.2}\footnote{Astrophysics Source Code Library identifiers: ascl:1601.005 and ascl:1110.007} (\citealt{knodlseder16}) to produce simulated event lists ({\it Level 3 Data}) containing sky arrival direction, energy,  arrival time, and other standard information for each event.
For each 20-minute block of a given field, we provided as input to {\sc ctobssim} an XML file containing the spectral, temporal and morphological information on the sources in the field and on the background components. Every observation was defined by a start and stop time (taken from the pointing plan), and source+background events were simulated in the whole energy range and FoV of the IRF. We always passed to {\sc ctobssim} different seeds for the random number generator and we assigned an observation identification (Obs.ID) to each 20-minute block.

We performed the simulations using Workflow Management Systems \&
Science Gateway technologies \citep{costa15} that allowed us to make use of high
performance computing capabilities with a customized interaction model
provided by the WP {``Simulation''}. We used  computational
resources  provided by INAF - Astrophysical Observatory of Catania in
the framework of the CHIPP Pilot Project\footnote{https://www.ict.inaf.it/computing/chipp/}. The used Linux cluster has a total of 192 cores (384 with Hyper-Threading), total RAM of 1024 Gb, 70 Tb storage and 3500 Gflops computing performance.

All simulated {scientific} data were then stored in an archive created by a dedicated WP (``Archive''). The main aim of the archive is to collect not only Level 3 data but also Level 4 (spectra, light-curves, sky images) and Level 5 (catalogues) data. The format of the stored data is  compliant with the {preliminary} CTA requirements. 

Finally, we performed a multi-wavelength association of the detected sources by means of an automatic identification pipeline. The pipeline first checks if there are known sources that can be potentially associated with the target, using positional coincidence and comparing their spectral and morphological properties. In case no viable association is found, the pipeline devises an optimized strategy for the multi-wavelength follow-up observations, quantifying the effort (telescopes, instruments, observing time) needed to properly cover an unidentified source. The present version of the pipeline incorporates an identification/strategy procedure only for the most abundant classes of Galactic VHE sources, PWNe and SNRs. In the future, we plan to extend this work and implement a dedicated strategy for the most numerous class of extragalactic VHE sources, namely Blazars.

\subsection{The ASTRI response functions}

The IRF of the ASTRI mini-array are needed   to simulate the observations as well as in the subsequent data analysis.

We used the IRF obtained through dedicated CTA Monte Carlo (MC) simulations of the Cherenkov light produced by extensive air showers previously simulated by {\it CORSIKA} \citep{heck98}. The {\it sim\_telarray} \citep{bernlor08} program (adopted by the CTA Consortium) and the official ASTRI pipeline (as part of the ASTRI Scientific Software) were adopted to derive the response of the ASTRI mini-array. In particular, the IRF was extracted from the CTA Prod3b\footnote{\url{http://www.cta-observatory.org/science/cta-performance/}} MC simulations ({\it Level 0 Data}), considering a mini-array of {nine SST} with a layout  {assumed at the site in Paranal (Chile)}, {as shown in Figure~\ref{layout_ASTRI}-left (mean distance between telescopes of $\sim$250 m)}. The incoming directions of the simulated showers are Zenith = 20 deg and {Azimuth at 0 deg and 180 deg}. The MC data have been reduced from Level 0 up to the IRF generation with A-SciSoft MC data reduction pipeline \citep{lombardi16}. {The gammaness (i.e. a gamma/hadron separation parameter) cuts for IRFs calculation were defined such as to have a 70\% efficiency for gammas in each estimated-energy bin (21 logarithmic bins between 0.012 TeV and 199.5 TeV) and in each off-axis bin (10 linear bins between 0 and 5 deg)}.

\begin{figure*}
   \centering
   \includegraphics[width=9.4cm]{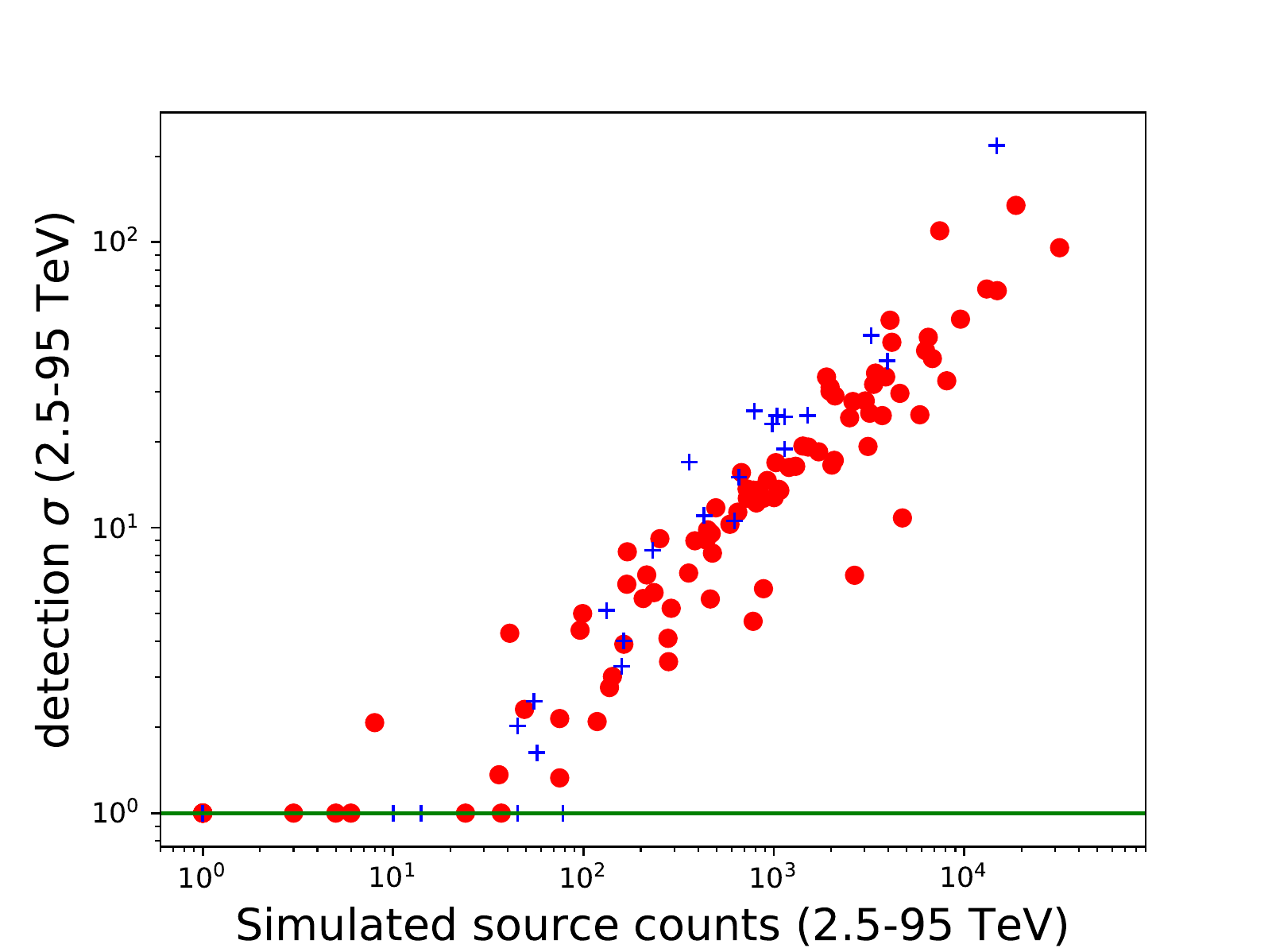}
   \includegraphics[width=9.4cm]{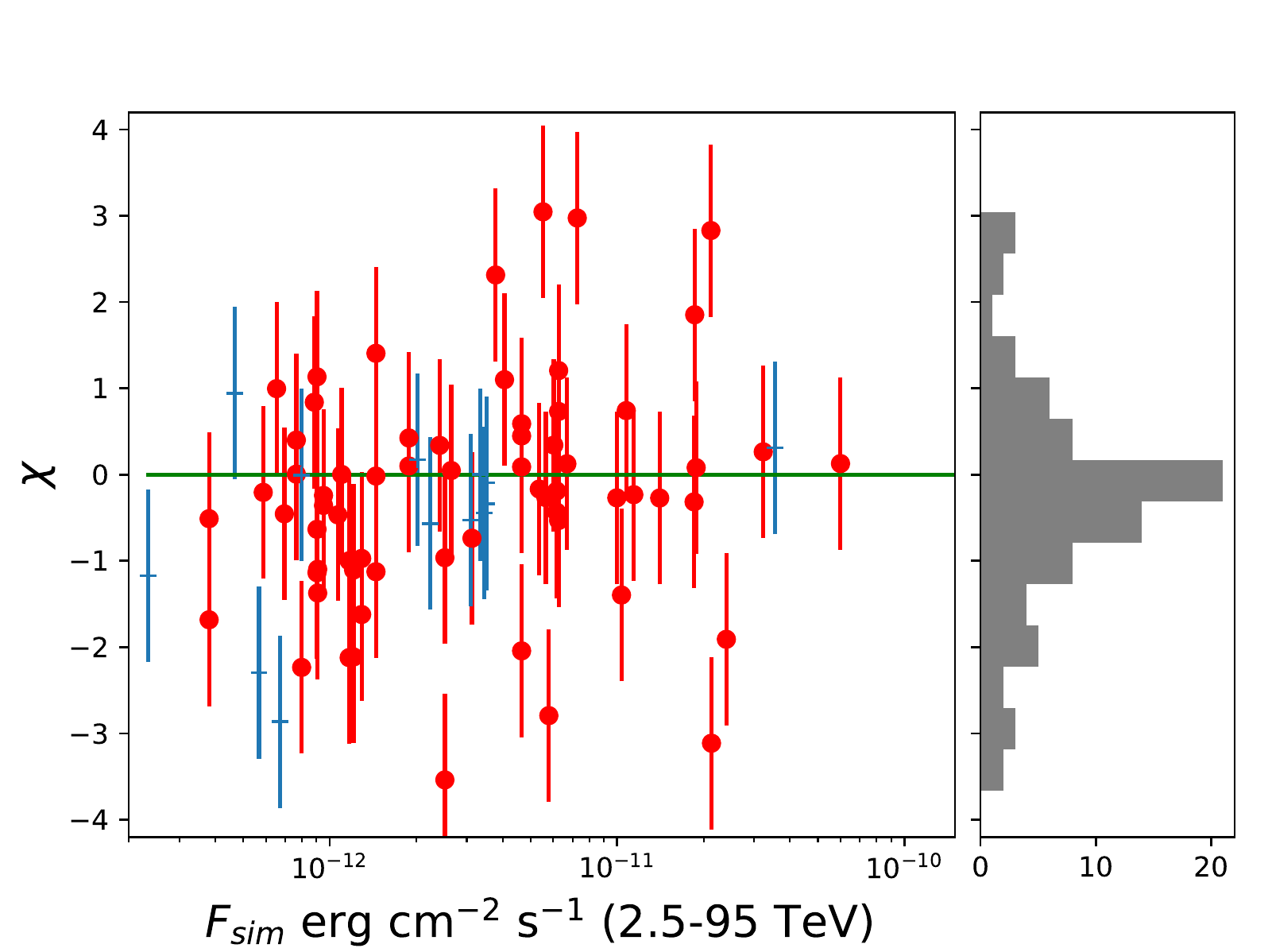}
      \caption{Left: detection significance (in units of standard deviation, $\sigma$, estimated as the square root of TS) as a function of the number of simulated source counts in the energy range 2.5--95 TeV; blue and red points represent point-like and extended sources, respectively, while the green line indicates the 1$\sigma$ level (for display purposes only, a number of undetected sources - i.e. with significance $\ll1\sigma$ - are shown to cluster on this line). Right: residuals, defined as $(F_{sim} - F_{obs})/(\Delta F_{obs})$, of the simulated versus observed fluxes, as a function of the 2.5--95 TeV  simulated flux, and their distribution (shown by the grey histogram); we show only the sources detected with a significance larger than $5\sigma$.}
         \label{Fig1}
\end{figure*}

The ASTRI mini-array IRF provides effective area, energy dispersion (Edisp) and point spread function (PSF) for the energy range 0.01--199.5 TeV (split into 21 logarithmic bins, as defined in \citealt{hassan17} and \citealt{acharyya19}) and for ten values of the OFF-axis angle (between 0 and 5 deg).
However, due to the energy threshold of the system of $\sim$1 TeV, the effective response of the ASTRI mini-array IRF is optimal in the energy range 1--200 TeV (see Fig.~\ref{layout_ASTRI}-right for a view of the ASTRI effective area).

{We simulated and analyzed the data considering only the IRF described above. This implies a fixed Zenith angle range (i.e. 0--20 deg) for all the nineteen fields, which is certainly not realistic. Taking into account the proper Zenith angle for each field would affect the results presented hereafter. However, quantifying this effect is beyond the scope of the paper and we will not discuss it further. }

\section{General results}

\subsection{Data analysis}\label{sec:genres}

The analysis of the data was performed with {{\sc ctools}-1.5.2}. We first carried out a systematic analysis of the whole data set  to derive the significance, mean flux and spectrum of each simulated source. For this preliminary step, we assumed to know the exact position and spatial morphology of each simulated source.

We performed a binned Maximum {Likelihood} analysis, rather than an un-binned one, because the latter was computationally extremely time-consuming and thus not convenient for a systematic analysis of such a  large data set. However, we checked {on four targets (two point-like and two extended)} used as test cases that the two approaches gave results {consistent within their uncertainties}.
Using {\sc ctbin} we created,  for each analyzed target,  a counts cube with 20 logarithmic bins in the energy range 2.5--95 TeV (where ASTRI is best calibrated) and  20$\times$20 spatial pixels of side 0.05 deg each, centered at the source position. The PSF, exposure, energy dispersion and background cubes were obtained using the tools {\sc ctpsfcube, ctexpcube, ctedispcube} and {\sc ctbkgcube}, respectively. For each of these tasks, we adopted the same energy selection of {\sc ctbin}, while for the energy dispersion and point spread function cubes, the spatial map was created considering 10$\times$10 pixels of side 1 deg, since these quantities vary slowly within the FoV and a finer binning is not necessary.

Subsequently, we determined the source spectral properties with {\sc ctlike}. In particular, we provide an input XML file with all the simulated sources (including their spatial and spectral properties), where we left free to vary only the spectral parameters for a single source of interest at a time (and hence, as mentioned above, assuming that its exact position and morphology are known). 
We used, for each source, a reference energy of 3 TeV for all the fitted spectral models.
For some sources with a complex spectrum, the data were simulated using a non-parametric spectral description (i.e. using tabular files). For example, this was the case for some extragalactic sources where the absorption from the extragalactic background light can be important. In such  cases, we left only a multiplicative normalization constant free to vary in the fit. 

\subsection{Source significance and validation of the results}

We analyzed {individually} each simulated field listed in Table 1. Therefore, since some of the 81 sources appear in more than one field, we investigated a total of 136 sources. Of these, we detected 83 sources (67 of which  seen only once) with a value of TS$\geq$25 (corresponding to a detection significance of $\sim5\sigma$). This number includes also three extragalactic sources (the AGN) and three of the serendipitous transients, which have 2.5--95 TeV fluxes in the range $0.4-4\times10^{-12}$ erg cm$^{-2}$ s$^{-1}$ and good agreement between simulated and observed spectral parameters. {All these were simulated with power-law spectral shapes ($\Gamma$ in the range $[1.4:2.6]$) and point-source morphology. {We obtained generally a good agreement} between the measured values and the simulated ones}. 
63 of the detected sources were spectrally modelled by power-laws, 14 by exponentially cut-off power-laws, 1 log-parabola, and {4 constant models (used as multiplicative constants, i.e. normalizations, for input cube map and file-function models)}. Instead, the number of extended detected sources is 68 (radial Gaussians: 45; elliptical Gaussian: 1; radial disk: 1; elliptical disk: 4; cube maps: 17).

The remaining simulated sources that were not detected were either too faint or poorly exposed and/or placed at very large off-axis angles resulting in small exposures. Amongst them, there are the dark matter candidates, {which were simulated with a 1--10 TeV integrated flux in the range $(3-11)\times10^{-17}$ erg cm$^{-2}$ s$^{-1}$.}

In Fig.~\ref{Fig1}-left, we present the source significance (in units of standard deviations, $\sigma$, estimated as the square root of TS) as a function of the number of simulated counts in the energy band 2.5--95 TeV. There is a good correlation between the two quantities, as it can be expected when the exposure times are similar (in our case within a factor of 2--3). {We note that these significances can be considered as upper limits on the values that can be obtained with the real data, where the sky model is not known {\it a priori}.}

For the sources detected with significance larger than 5$\sigma$, we compared the observed ($F_{obs}$) and simulated ($F_{sim}$)  fluxes in the  2.5--95 TeV range, by computing for each  source the residuals $\chi$=$(F_{sim}-F_{obs})/\Delta F_{obs}$, where $\Delta F_{obs}$ is the 1$\sigma$ uncertainty of $F_{obs}$.    Fig.~\ref{Fig1}-right shows that these residuals, as expected, are independent on the source flux and that there is a good agreement between $F_{sim}$ and $F_{obs}$. However, the distribution of the residuals {does not properly follow the shape of a Gaussian distribution, as it would be instead expected}. 
{We explain this with the presence of a {\it global} systematic uncertainty, possibly due to the uniform choices adopted for the analysis of all the sources, which is not optimized for specific cases, {or due to the poor statistics in some energy bins that makes the flux estimation non-gaussian,} or fixing the parameters of the non-target sources in the FoV to the simulation input values, which may not match exactly the simulated data. We evaluate the {\it global} systematic uncertainty for the reconstructed, integrated fluxes to be of the order of $\sim8\%$: this allows us to obtain a statistically acceptable fit to a constant of the integrated fluxes shown in Figure~\ref{Fig1}-right}.

\section{Individual cases}

In this Section we report a few examples of a more detailed analysis of representative sources, in order to better illustrate the results of the simulations and the expected performance of the ASTRI mini-array.  
 These examples  provide a variety of different morphological and spectral shapes as well as temporal variability, unlike our simulated extragalactic sources. In addition, these test-case sources are located in crowded fields, where the analysis is more complex. We note that a thorough assessment of the scientific capabilities of the mini-array in the case of specific sources would require more extensive simulations of each target that is outside the scope of a data challenge.
 
\subsection{The Crab nebula and pulsar}

The Crab is one of the brightest and best studied TeV $\gamma$-ray sources. It has been also traditionally used as a calibration source for high-energy instruments, although this should be reconsidered at GeV energies in the light of the variability observed by AGILE and Fermi \citep{tavani11,fermi11}. It was the first  $\gamma$-ray source detected by a Cherenkov telescope \citep{weekes89} and it is currently regularly monitored at TeV energies by MAGIC, H.E.S.S., VERITAS and HAWC \citep{aleksic15,holler17,meagher15,abeysekara17} and FACT\footnote{\url{https://fact-project.org/monitoring/index.php?y=2012&m=12&d=13&source=5&timebin=12&plot=all}} \citep{anderhub13}. The angular size of the Crab nebula  decreases with energy and in the TeV band it is expected to be $<0.2$ deg \citep{aharonian00}. Very recently, its TeV extension has been measured with H.E.S.S. ($52.2'' \pm 2.9'' \pm 7.8''_{sys}$; \citealt{holler17}). 

We simulated $\sim$100 hr of observations of the Crab (field 18),  that was modeled taking into account the emission from both the  pulsar and the  nebula. Although we do not expect to detect the Crab pulsar in this short  observation (less than 20 expected counts), we included it in the sky model with a power-law spectrum \citep{ansoldi16}. 
Because the on-axis ASTRI mini-array PSF, in the 1--100 TeV, spans the range $0.06-0.1$ deg (hence larger than its measured extension), the Crab nebula was simulated as a point-like source with a log-parabola model \citep{aleksic15} of the form: 
\begin{equation}
f(E) = 1.89\times10^{-12}  \left(\cfrac{E}{\text{3 TeV}}\right)^{-\alpha -\beta\text{ }ln(E/\text{3 TeV})} \cfrac{\text{ph.}}{\text{cm}^{2} \text{ s} \text{ TeV}} 
\end{equation}
with $\alpha=2.698$ and $\beta = 0.104$.

\begin{figure}
   \centering
   \includegraphics[width=\hsize]{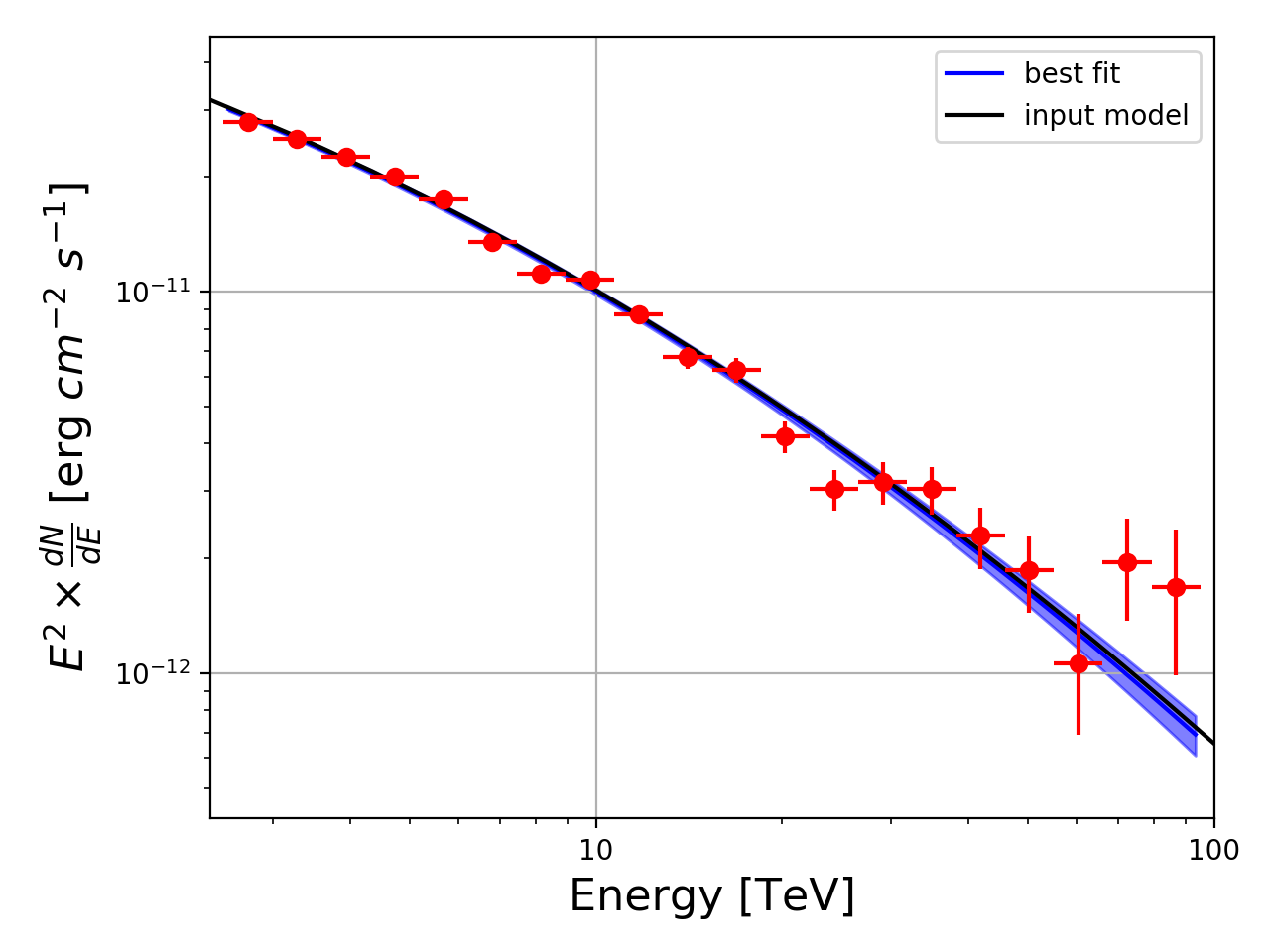}
      \caption{The spectrum of the Crab in $\text{E}^2f(\text{E})$, in the energy range 2.5--95 TeV, where the red points (all significative at $3\sigma$) are the spectral fluxes calculated with {\sc csspec}, the black line is the simulated model, and the blue shadow is the butterfly of the best-fit model calculated with {\sc ctbutterfly}. A good agreement between simulated and observed data is found.}
         \label{Fig4}
\end{figure}

Our spectral analysis yielded the following best-fit parameters for the nebula: $K=(1.87\pm0.02)\times10^{-12}$ photons cm$^{-2}$ s$^{-1}$ TeV$^{-1}$, $\alpha=2.68\pm0.02$ and $\beta=0.115\pm0.015$, which are consistent within $1\sigma$ with the expected ones, as shown in Fig.~\ref{Fig4}. For the sake of completeness, we tried also to fit the Crab PSR spectrum, fixing the Crab nebula spectral parameters to those reported above. We found that the Crab PSR component is indeed not significant. We did not perform any timing analysis because the number of pulsar events was too low (14 counts).
   
\begin{figure*}
 \centering
\includegraphics[width=9.8cm]{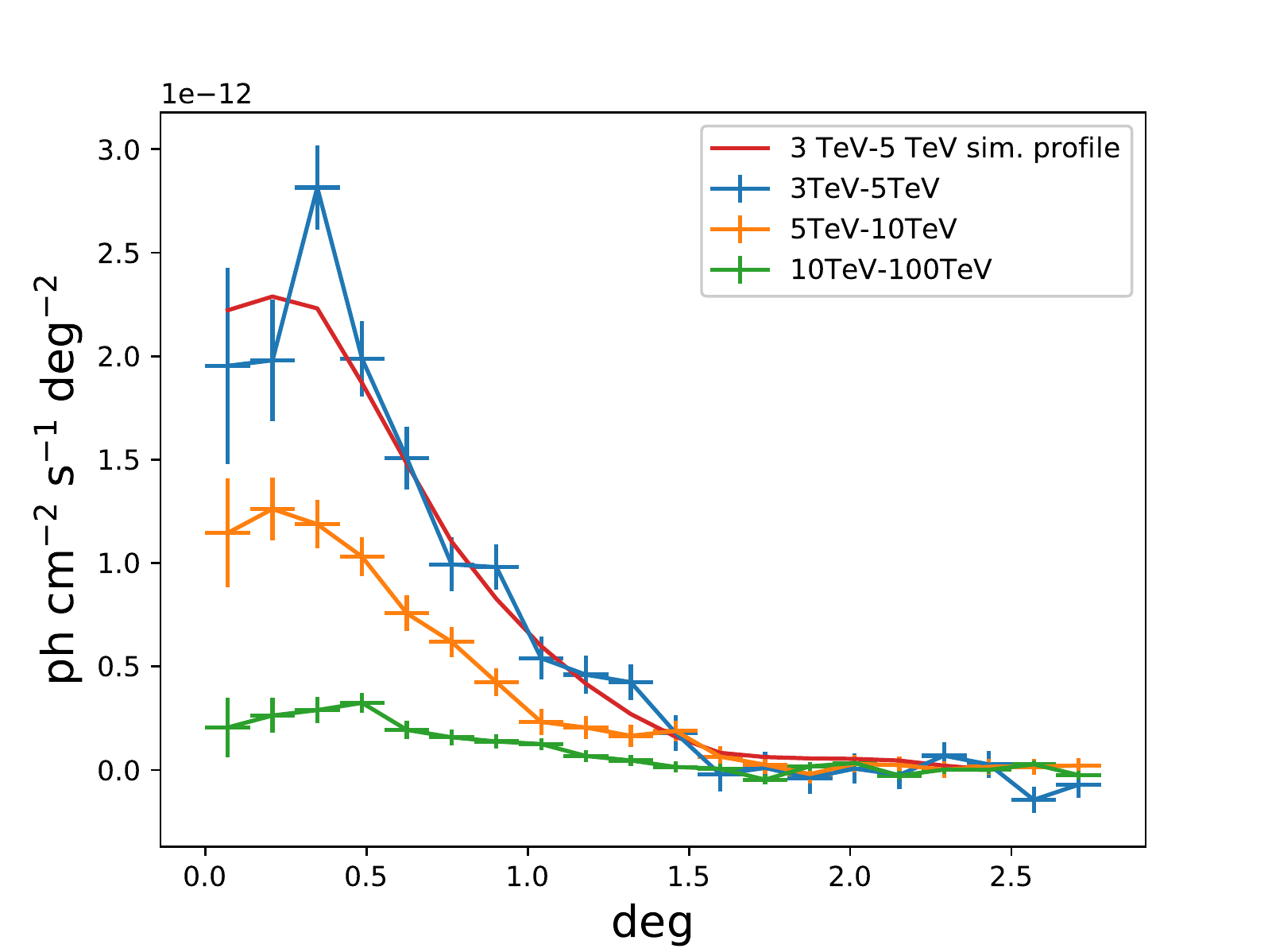}
\includegraphics[width=9.1cm]{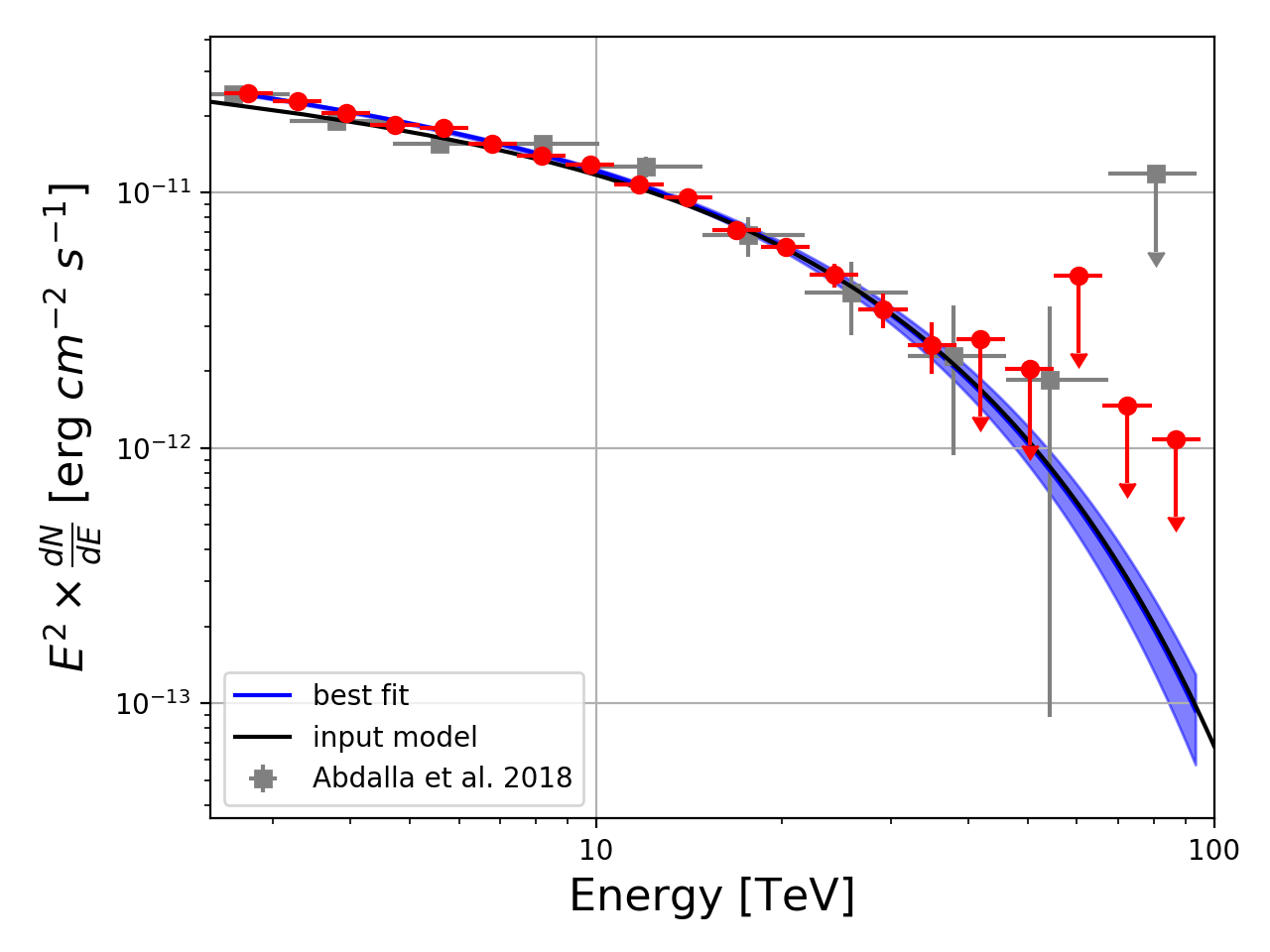}
      \caption{Left: background-subtracted radial profile of HESS J1825--137, {compared with the simulated 3--5 TeV radial profile (red solid line)}. Right: mean spectrum (red points), simulated model (black line), best-fit model (the blue butterfly) obtained for HESS J1825--137. {We added for comparison also the spectrum reported in \citet{abdalla18b} (grey points) in their Figure 2-right.}}
         \label{fig:lshessj1825}
\end{figure*}
These results indicate that the ASTRI mini-array, in 100 hr, will be able to extend the maximum detection energy of the Crab nebula up to $\sim100$ TeV, with   a significance of the spectral points larger than $3\sigma$. This is an important improvement with respect to the current generation of Cherenkov telescopes:  VERITAS detected the Crab up to a maximum energy of $\sim30-40$ TeV in $\sim115$ hr \citep[e.g.][]{meagher15}, MAGIC and H.E.S.S. up to $\sim10-20$ TeV in $\sim360$ hr and $\sim10$ hr, respectively \citep{aleksic15,ansoldi16,holler17}. We also note that recently HAWC, the {\it Tibet Air Shower Array} {and MAGIC} measured the first Crab nebula photons at energies $\geq100$ TeV in $\sim837$, $\sim719$ observing days and $\sim 50$hr, respectively (\citealt{abeysekara19, amenomori19};\footnote{\url{https://www.icrc2019.org/uploads/1/1/9/0/119067782/vlza_crab_peresano_icrc2019.pdf}; \url{https://www.icrc2019.org/uploads/1/1/9/0/119067782/vlza_crab_peresano_icrc2019.pdf}}).

\subsection{The bright Pulsar Wind Nebula HESS J1825--137}
\label{hessj1825_sec}
HESS J1825--137 is one of the brightest extended sources in the TeV sky. It is spatially coincident with the pulsar PSR B1823--13 that is most likely powering it \citep[e.g.][]{pavlov08}. The source distance is estimated to be about 4 kpc \citep{cordes02}. HESS J1825--137 shows a multi-wavelength extended emission detected in the soft X-ray energy band (with a size of $\sim5'$; e.g. \citealt{finley96, gaensler03}), in the radio band \citep[e.g.][]{castelletti12}, and in the $\gamma$-ray by FERMI (with an extension of $\sim1.1$\textdegree; \citealt{ackermann17}) and HAWC \citep{abeysekara17b}. The source has been extensively observed by H.E.S.S. in the last decade \citep[e.g.][and reference therein]{abdalla19}. These observations confirmed the existence of  diffuse TeV emission which extends for approximately 1.5 deg from the pulsar position \citep{abdalla19}, mainly towards the south-east direction \citep[e.g.][]{Aharonian06b}. HESS J1825--137 is characterized by a clearly energy-dependent morphology, with a spectrum becoming softer with increasing distance from the pulsar \citep{Aharonian06b,abdalla19}. This can be explained as the effect of  radiative cooling of the electrons, or by changes of the injection spectrum of electrons during the source history, or by diffusion or advection effects. Its average spectrum, extracted from a region of 0.8 deg, can be modelled better with a cut-off power-law rather than a simple power-law \citep{abdalla19}.

HESS J1825--237 is in the same FoV of the $\gamma$-ray binary LS~5039, and our simulated total exposure time for the source is 300 hr, coming from fields 01 and 02 in Table 1 (in field 17 the source was too far off axis). We considered the same spatially varying spectrum found by \citet{abdalla19} (their figure 8), although we did not consider the variation of the PWN size with energy. We also extrapolated it, following the relations showed in figure 9 of \citet{abdalla19}, to slightly larger distances (about 10-20 arcminutes) in the south-east and in the north-west directions in order to avoid a sharp cut of the source morphology. 
However, for the aim of this paper, we only estimated the HESS J1825--137 background-subtracted radial profile (Fig.~\ref{fig:lshessj1825}-left) along the south-east direction using the green region shown in Fig.~\ref{fig:ls5039_ima}-right,     similar to the one adopted in \citet{Aharonian06b}.  We estimated the background from the closest blank field of the data challenge. 
Our results show that HESS J1825--137 stands-out above the background, with a significance larger than 3$\sigma$, up to a distance of $\sim1.5$ deg from the pulsar position. This demonstrates the capabilities of the ASTRI mini-array to detect very extended sources.

We  estimated the total spectrum extracted from a circular region of 0.8 deg (as presented by \citealt{abdalla19}) centered at $RA = 276.40$ deg and $DEC = -13.967$ deg. In particular, we used a disc morphology with fixed parameters and we fitted the data with both a power-law and a cut-off power-law model. As expected, we found that the latter provides a better fit than the power-law (TS = 13908 vs TS = 13745, for 1 additional d.o.f.) indicating that the cut-off is significantly detected. The best-fit parameters (shown in Fig.~\ref{fig:lshessj1825}-right) of the exponential cut-off power-law are $N_0 = (1.90\pm0.04) \times 10^{-12}$ ph cm$^{-2}$ s$^{-1}$ TeV$^{-1}$, $\Gamma=2.24 \pm 0.04$ and $E_{cut} = 19\pm2$ TeV (for a reference energy of 3 TeV).
Such values are totally in line with those reported in \citet{abdalla19} in $\sim400$ hr of exposure time. Thus,   the ASTRI mini-array will not only provide a better spatial resolution (allowing a more detailed study of the energy-dependent morphology) but it also may provide comparable, or even significantly reduced, uncertainties of the spectral parameters, for the same exposure time as H.E.S.S..

\subsection{The binary LS 5039}
\label{LS-5039}
LS~5039 is a $\gamma$-ray binary composed of a compact object and an ON6.5V(f) star \citep{Clark2001, McSwain2004},
in an eccentric, 3.9~day orbit (e=0.31$\pm{0.04}$). It was detected above 250\,GeV  for the first time with H.E.S.S.
\citep{Aharonian2005} and its orbital modulation was the first one observed at TeV energies \citep{Aharonian2006}.
Its  0.2--10\,TeV time-averaged luminosity is 7.8$\times$10$^{33}$~erg~s$^{-1}$, for a  distance of 2.5$\pm{0.1}$~kpc \citep{Casares2005}.

Orbital phase-resolved spectroscopy showed the existence of two different spectral states: a high-state in the orbital phase range  0.45$< \phi <$0.9, comprising the inferior conjunction (the apastron passage is at $\phi$=0.5), and a low-state in the remaining part of the orbit \citep{Aharonian2006}.
Adopting a power-law model with exponential cutoff ($dN/dE \sim E^{-\Gamma}\, \exp(-E/E_{cut})$), the 
0.2--10\,TeV high-state spectrum gives $\Gamma=1.85\pm0.06_{\mathrm{stat}} \pm 0.1_{\mathrm{syst}}$
and  $E_{cut}=8.7\pm2.0$~TeV, while a simple steeper power-law is  a good description of the low-state spectrum ($\Gamma=2.53\pm0.07_{\mathrm{stat}} \pm 0.1_{\mathrm{syst}}$). 
The 0.2-10\,TeV luminosity is 1.1$\times$10$^{34}$~erg~s$^{-1}$  and 4.2$\times$10$^{33}$~erg~s$^{-1}$ in the high and low state, respectively \citep{Aharonian2006}.

As in most of the  $\gamma$-ray binaries, the nature of the compact object in LS 5039 is still unknown: both the microquasar (a black hole in accretion from the massive stellar wind)  and the young non-accreting pulsar scenarios are possible \citep{Dubus2013}.
The periodic variability of the shape of its radio emission {may originate from} the interaction of a pulsar wind with the wind of the early type companion star \citep{Moldon2012}, contrary to the first interpretation of the resolved radio emission as a jet in a microquasar \citep{Paredes2000}.

\begin{figure}
 \centering
   \includegraphics[width=\hsize]{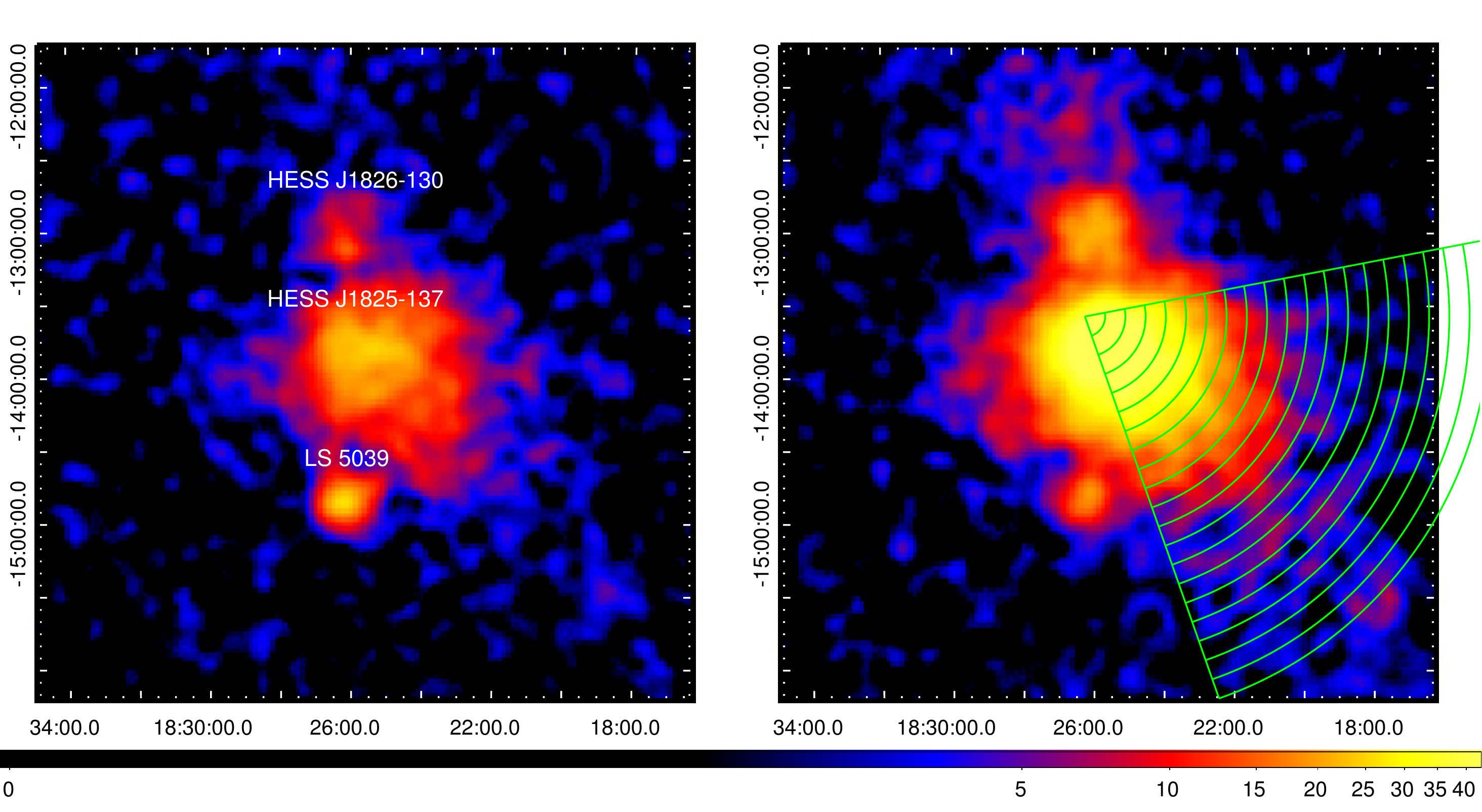}
      \caption{Background subtracted 2.5--95\,TeV sky-maps (in counts per pixel; pixel size of 0.03 deg) as obtained for the high state data (50\,h, left panel) and low state data (250\,h, right panel) of LS 5039, in celestial coordinates. The nearby PWN HESS~J1825$-$137, visible in the map, is discussed in section~\ref{hessj1825_sec}, where the green regions are used to estimate its radial profile. The latter is centered on the position of the pulsar PSR B1823-13.}
         \label{fig:ls5039_ima}
\end{figure}

\begin{figure}
 \centering
   \includegraphics[width=\hsize]{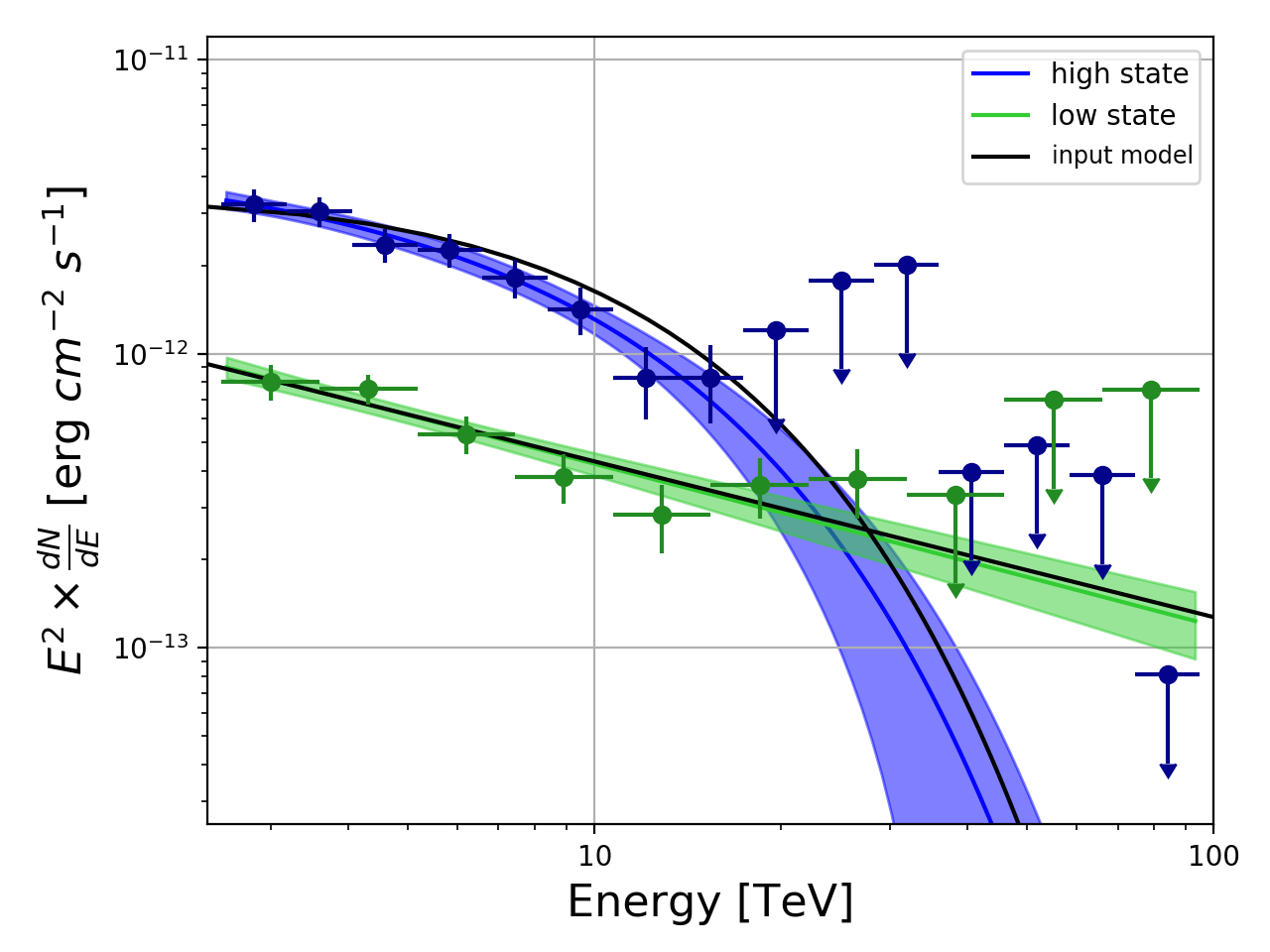}
      \caption{LS~5039 simulated spectra for the high state (50\,h, blue) and low state (250\,h, green). The black line is the simulated model,  data points are the spectral fluxes calculated with {\sc csspec}, while the blue/green line and shaded area are the best-fit model and relative butterfly, respectively, as obtained with {\sc ctlike}.}
         \label{fig:ls5039}
\end{figure}

We simulated observations of 50\,h for the high state (field 01) and 250\,h for the low state (field 02), using the spectra described above. The source is too far off-axis in the field 17 and not considered here.
Fig.~\ref{fig:ls5039_ima} shows the sky-maps in the 2.5--95\,TeV range for the two states, while Fig.~\ref{fig:ls5039} shows the spectra and best-fit models. \\
\indent
In the high spectral state (in blue in Fig.~\ref{fig:ls5039}), the source is detected with a significance of $\sim$25 $\sigma$ (TS=600.5, 2.5--95\,TeV) and the best fit  parameters are $\Gamma=2.1\pm0.3$, $E_{cut}=8.7\pm4$~TeV, 
and normalization (flux at 1 TeV)  N=(2.9$\pm$0.6)$\times 10^{-12}$ photons cm$^{-2}$ s$^{-1}$ TeV$^{-1}$ (to be compared with the simulated value of  2.28$\times 10^{-12}$ photons cm$^{-2}$ s$^{-1}$ TeV$^{-1}$). 
The best fit parameters  agree  with those of the model within 2\,$\sigma$, although the presence of the cut-off at $\sim$9\,TeV, together with the analysis truncated at 2.5\,TeV, leads to large statistical uncertainties.
We verified that the detection of the spectral cutoff is statistically significant in the high state: indeed, a fit with a single power-law gives $\Gamma=2.8\pm0.1$, but it is worse by 3.7\,$\sigma$  than that with a cutoff power-law.\\
\indent
In the low spectral state (in green in Fig.~\ref{fig:ls5039}),  the source is detected with a significance of $\sim$18 $\sigma$
and we obtained  $\Gamma=2.55\pm0.08$ and N=0.9$\pm$0.1 $\times 10^{-12}$ photons cm$^{-2}$ s$^{-1}$ TeV$^{-1}$ (input 0.91 $\times 10^{-12}$ photons cm$^{-2}$ s$^{-1}$ TeV$^{-1}$). 
Unlike the high-state data, in the low-state spectrum the simulated ASTRI mini-array data alone do not constrain a possible high-energy cut-off. {A constraint could be obtained assuming an `a priori' knowledge of the photon index. For example, freezing the power-law parameters to the low-state H.E.S.S. values ($\Gamma=2.53$ and N$_{1\text{TeV}}$=0.91 $\times 10^{-12}$ photons cm$^{-2}$ s$^{-1}$ TeV$^{-1}$), we would obtain a 3\,$\sigma$ lower limit on the cutoff of $E_{cut}>46$\,TeV.}
This value, well within the ASTRI mini-array energy range, is above the H.E.S.S. coverage and is consistent with a scenario where the energy of the spectral cutoff increases from the high state ($E_{cut}\sim 9$\,TeV) to the low state ($>45$\,TeV).

The short-term as well as long-term variability of the orbital TeV emission of LS~5039 and relative spectral cut-off is worth investigating both within the forthcoming ASTRI  mini-array observations as well as with respect to previous H.E.S.S. data. Hence it is vital to recover the very low energy of the IRF, below 2.5\,TeV, stabilizing the spectrum with lower energy photons, where the highest statistics resides.

\subsection{The P3 binary in the Large Magellanic Cloud}

LMC P3 is the only extragalactic $\gamma$-ray binary known to date. It has an orbital period of 10.3 days and is associated with a massive O5III star located in the Large Magellanic Cloud (LMC) supernova remnant DEM L241 \citep{Corbet+16}. Both its X-ray and radio emissions are modulated at the orbital period, but are in anti-phase with the $\gamma$-ray modulation. These results, together with the high ratio between the $\gamma$-ray and the X-ray flux, lead to the identification of this source as a high-mass $\gamma$-ray binary \citep{Corbet+16}. At $\gamma$-ray energies LMC P3 is significantly more luminous than any other $\gamma$-ray binary, since its 0.2--100 GeV luminosity is L = 2.5$\times 10^{36}$ erg s$^{-1}$  \citep{Ackermann+16}. Moreover, it is at least 10 times more luminous in radio and X-rays than LS 5039 and 1FGL J1018.6-5856 \citep{Corbet+16}. This is very peculiar, since both the luminosity of the companion star and the orbital separation of LMC P3 are comparable to those of these two binary systems. 

The extensive coverage of the LMC carried out with H.E.S.S. since 2004 (effective exposure $\sim$100 hr) showed that the TeV emission from LMC P3 has an orbital variability \citep{HESScollaboration+18}. If the orbital phase $\phi$ = 0 is defined as the maximum of the GeV light curve at MJD 57410.25, LMC P3 was clearly detected as a point-like source (at a 6.9$\sigma$ confidence level) only in the orbital phase range between 0.2 and 0.4.
Its time-averaged spectrum (accumulated along the whole orbit) is a power-law  with  $\Gamma$ = 2.5 $\pm$ 0.2 and  normalization N = (2.0 $\pm$ 0.4)$\times 10^{-13}$ ph cm$^{-2}$ s$^{-1}$ TeV$^{-1}$ (at 1 TeV). On the other hand, for the spectrum of the on-peak phase only, the photon index decreases to $\Gamma$ = 2.1 $\pm$ 0.2 and N = (5$\pm$1)$\times 10^{-13}$ ph cm$^{-2}$ s$^{-1}$ TeV$^{-1}$ (at 1 TeV). The flux above 1 TeV varies by a factor larger than 5 between on-peak and off-peak parts of the orbit, since it is (5 $\pm$ 2)$\times 10^{-13}$ and $<$ 0.88 $\times 10^{-13}$ ph cm$^{-2}$ s$^{-1}$, respectively.

Spectroscopic observations of this source \citep{vanSoelen+19} were performed using the High Resolution spectrograph (HRS) of the Southern African Large Telescope (SALT). They revealed that the binary orbit is slightly eccentric ($e$ = 0.40 $\pm$ 0.07). The phase of the superior conjunction is $\phi$ = 0.98, close to the maximum in the Fermi-LAT light curve, while inferior conjunction is at phase $\phi$ = 0.24. This orientation may explain the anti-phase between the GeV and TeV light curves. The mass function is $f$ = 0.0010 $\pm$ 0.0004 $M_{\odot}$, which favours a neutron star compact object.

\begin{figure}
 \centering
   \includegraphics[width=\hsize]{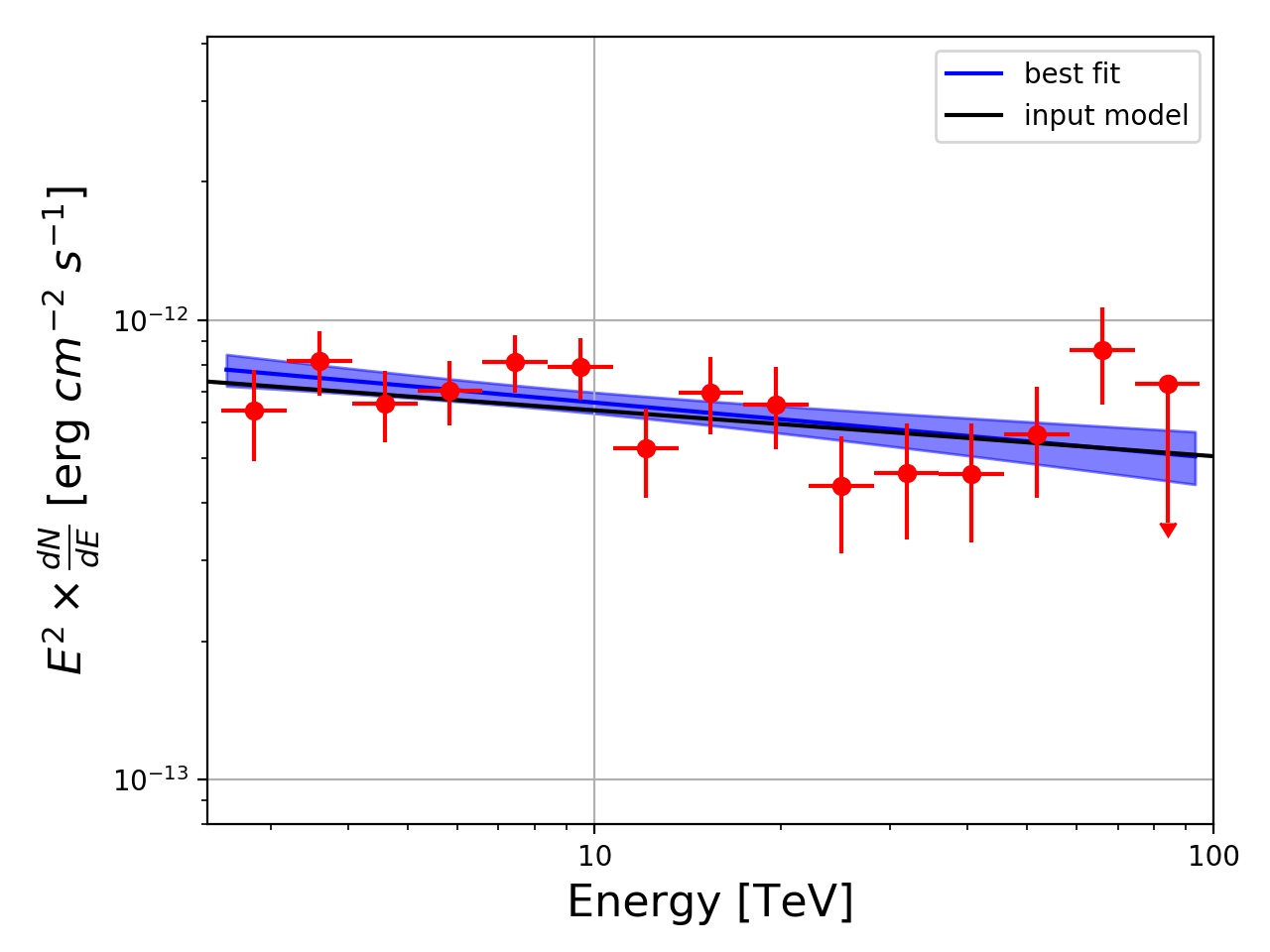}
      \caption{The spectral energy distribution of LMC P3, in the energy range 2.5--95 TeV, where the red points are the   fluxes calculated with {\sc csspec}, the black line is the simulated model, and the blue shadow is the butterfly of the best-fit model.}
         \label{lmcp3}
\end{figure}

We simulated  200 hr of time-constrained  observations in order to observe the source only during the orbital phase range $\phi$ = 0.2--0.4 (field 10). In the simulated data, LMC P3 is detected with a significance of TS $>$ 500 (i.e. $\geq22\sigma$). The best-fit power-law parameters are $\Gamma$ = 2.12 $\pm$ 0.05 and  N = (5.4 $\pm$ 0.6)$\times 10^{-13}$ ph cm$^{-2}$ s$^{-1}$ TeV$^{-1}$ (at 1 TeV), fully consistent with the simulated values. As shown in Fig.~\ref{lmcp3}, the source is clearly detected up to $\sim$70 TeV. This is well above the high-energy limit of 20 TeV reached with H.E.S.S. in 100 hr.

\subsection{The supernova remnant  W28}

We simulated a set of ASTRI mini-array observations of the sky region around the Galactic coordinates l,b $\sim$ (5.5,-0.5)  for a total exposure of about 300 hours (field 12). 
This field covers about 7 degrees of the inner Galactic plane and it is rich of known TeV sources, among which the supernova remnant W28 is one of  the  most prominent.   
W28 is an old SNR ($2-3.5\times10^4$ yrs)  interacting with a complex of massive molecular clouds with masses of $\sim10^4$ M$_{\odot}$. 
It is a source of an intense flux of $\gamma$-rays, extending  from $\sim$100 MeV to $\sim$10 TeV, with a soft power-law spectrum (photon index of $\sim$ 2.7; \citealt{aharonian08,giuliani10}).
The $\gamma$-ray emission  is believed to arise from the interaction of the gas inside the molecular clouds and the hadronic cosmic rays accelerated by the shock of the SNR.   
H.E.S.S. did not detect any sign of a cut-off in the spectrum of this object. 
The detection of a drop of the spectrum at very high energy can test if this SNR was a PeVatron in its early life. 
In fact, if W28 accelerated particles up E$ \geq 10^{15}$ eV in the past (and if the ambient medium is able to confine them) the $\gamma$-ray spectrum should continue without a cutoff up to, at least, 100 TeV.   
A cut-off in the spectrum can instead have two explanations: either  this accelerator  never reached such a high energy and/or  the  propagation of cosmic ray changes regime (i.e. from diffusion to free streaming) above a certain energy.
We simulated the W28 spectrum with a power-law without cut-off (i.e. assuming a PeVatron-like spectrum) and we tested the ASTRI mini-array capability to derive a lower limit on the maximum energy of the  accelerated protons.
In our analysis, we derived the spectrum of this source using {\sc csspec}. The resulting spectral points are significant (more than 3$\sigma$) up to about {80} TeV.

\begin{figure}
   \centering
   \includegraphics[width=8.3cm]{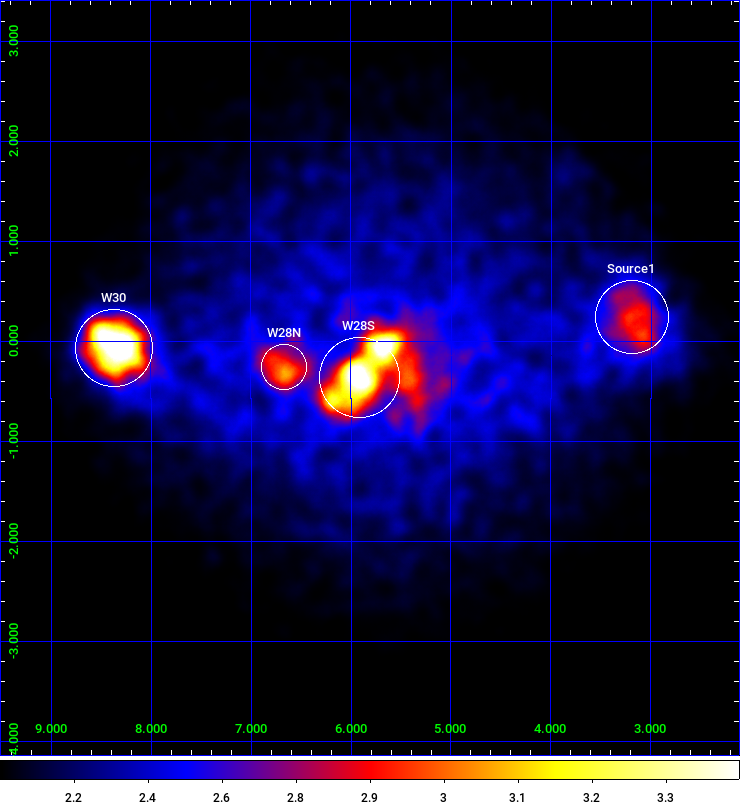}
      \caption{The simulated W28 field in Galactic coordinates.} 
               \label{w28a}
\end{figure}

\begin{figure}
   \centering
   \includegraphics[width=\hsize]{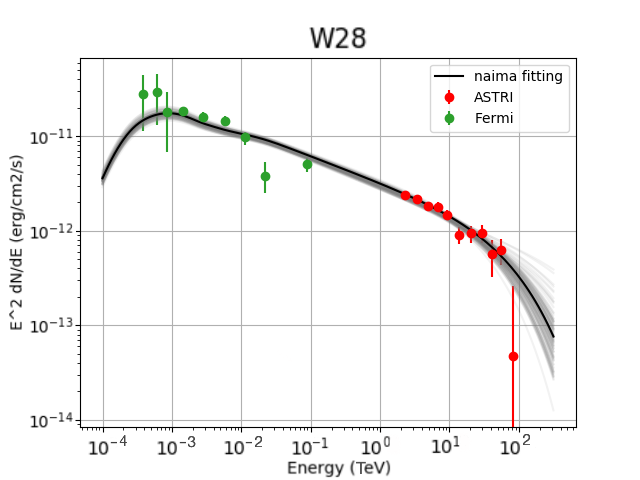}
      \caption{The spectrum of W28 combining Fermi data and ASTRI simulations. }
                \label{w28b}
\end{figure}

Similar to  what was found in other analysis of this class of sources \citep{aharonian08}, a single power-law can connect the GeV spectral data points with the TeV spectrum.
We then used the Fermi and ASTRI data to better constrain the spectral parameters of the $\gamma$-ray emitting protons.  
We used the Naima package \citep{naima} to fit the spectrum from few hundreds of MeV to about 100 TeV.  
The results are shown in Fig.~\ref{w28b}.

{As a result of this fit, a cut-off in the protons distribution with energy less than 300 TeV can be excluded by these set of data at 90$\%$ of c.l. }
This shows how this kind of analysis can lead to observe the cosmic rays at an energy very near the knee in the  spectrum and then to constrain the role of SNR in the acceleration of Galactic cosmic rays.

\noindent

\section{Conclusions}

{\it ACDC} is an {\it INAF} project aiming at simulating realistic end-to-end operations of three years of observations of the future ASTRI mini-array. We considered a layout of the mini-array {for an assumed site in Paranal}. 
We focused on the fields of 20 reference targets of different types of objects, chosen for their scientific relevance in the TeV sky. Thanks to a realistic pointing plan, we simulated up to a total of {5160 hr} of useful time over three years. In addition, the large FoV ($\sim10$\textdegree) of the ASTRI mini-array allowed us to include in the simulation also a large number of H.E.S.S. and Fermi sources falling into the fields of the reference targets, for a total of 81 simulated point-like and extended objects.

In this paper, we have shown the results of a systematic analysis with {{\sc ctools}-1.5.2} of the ASTRI mini-array data-challenge. We found that 67 sources  were detected with a significance larger than $5\sigma$, with a good agreement between their observed {and simulated} spectral parameters. We also estimated a global systematic error, which includes both uncertainties in the IRF and limits of the analysis, of $\sim8\%$ on the estimated fluxes.

Our results show that the ASTRI mini-array will be able to carry {out} scientific observations of the $\gamma$-ray sky up to about 100 TeV, extending significantly the range of energies covered by current Cherenkov instruments. We showed that the ASTRI mini-array will permit to investigate the properties of a large number of different types of objects, offering the possibility to study the morphology of extended sources (thanks to its high angular resolution of $\sim0.05-0.15$ deg), to detect faint TeV objects and also to discover serendipitous transient sources.
The findings presented in this work show that the capabilities of the ASTRI mini-array will provide results comparable, or even better than, those currently available with H.E.S.S. {at high energies}. 
We finally note that the ASTRI mini-array will be important in the search of PeVatron candidates.

\section*{Acknowledgements} 

We would like to thank the anonymous reviewer, who provided useful suggestions for improving the final manuscript.

The authors acknowledge contribution from the grant INAF CTA (PI: P. Caraveo). 

This work is supported by the Italian Ministry of Education, University, and Research (MIUR) with funds specifically assigned to the Italian National Institute of Astrophysics (INAF) for the Cherenkov Telescope Array (CTA), and by the Italian Ministry of Economic Development (MISE) within the âAstronomia Industrialeâ program. 

{This research has made use of the CTA Montecarlo production (prod3b) provided by the CTA Consortium and Observatory.}

This research made use of {\sc ctools}, a community-developed analysis package for Imaging Air Cherenkov Telescope data. {\sc ctools} is based on GammaLib, a community-developed toolbox for the scientific analysis of astronomical gamma-ray data.


\bibliographystyle{aa} 
\bibliography{biblio}

\end{document}